\documentclass{IEEEtran}
\usepackage[utf8]{inputenc}

\usepackage{graphicx}
\usepackage{epstopdf}
\usepackage{enumitem}
\usepackage{listings}
\usepackage{xcolor}
\usepackage{algorithm} 
\usepackage{algpseudocode}
\usepackage{todonotes}
\usepackage{csquotes}

\usepackage{booktabs}    
\usepackage{tabularx}
\usepackage{multirow}
\usepackage{booktabs}
\usepackage{siunitx}
\usepackage{url}
\usepackage{nicefrac}
\usepackage{float}
\usepackage{tikz}
\usetikzlibrary{patterns}
\usepackage{colortbl}
\usepackage{pgfplots}
\usepackage{pgfplotstable}
\usepackage{pgf-pie,etoolbox}

\usepackage{subcaption}

\usepackage{amsmath}
\usepackage{amsthm}
\usepackage{amssymb}
\usepackage{cleveref}

\usepackage{pifont}

\definecolor{my_orange}{RGB}{255, 102, 0}
\definecolor{my_green}{RGB}{0, 128, 0}
\definecolor{my_cyan}{RGB}{0, 255, 255}
\definecolor{my_yellow}{RGB}{255, 255, 0}
\definecolor{revcolor}{rgb}{1,0.5,0}

\newcommand\rev[1]{{\color{black}#1}}

\pgfplotsset{compat=1.18}

\usepackage[acronym]{glossaries}
\glsdisablehyper
\newacronym{hbm}{HBM}{high-bandwidth memory}
\newacronym{hls}{HLS}{high-level synthesis}
\newacronym{fpga}{FPGA}{field programmable gate array}
\newacronym{asic}{ASIC}{application-specific integrated circuit}
\newacronym{it}{IT}{information technology}
\newacronym{ssd}{SSD}{solid-state disk}
\newacronym{hdd}{HDD}{hard-disk drive}
\newacronym{sram}{SRAM}{static random access memory}
\newacronym{clb}{CLB}{configurable logic block}
\newacronym{alm}{ALM}{adaptive logic module}
\newacronym{iob}{IOB}{input-output block}
\newacronym{ioe}{IOE}{input-output element}
\newacronym{sll}{SLL}{super-long-line}
\newacronym{sm}{SM}{switch matrix}
\newacronym{dram}{DRAM}{dynamic random access memory}
\newacronym{cpu}{CPU}{central processing unit}
\newacronym{gpu}{GPU}{graphic processing unit}
\newacronym{hdl}{HDL}{hardware description language}
\newacronym{cmos}{CMOS}{complementary metal-oxide semiconductor}
\newacronym{flop}{flop}{floating-point operation}
\newacronym{pdu}{PDU}{power distribution unit}
\newacronym{pue}{PUE}{power usage effectiveness}
\newacronym{lca}{LCA}{life cycle assessment}
\newacronym{uk}{UK}{united-kingdom}
\newacronym{dvs}{DVS}{dynamic voltage scaling}
\newacronym{avs}{AVS}{adaptive voltage scaling}
\newacronym{dfs}{DFS}{dynamic frequency scaling}
\newacronym{ldmc}{LDMC}{logic delay measurement circuit}
\newacronym{pmu}{PMU}{power manager unit}
\newacronym{pcb}{PCB}{printed circuit board}
\newacronym{ram}{RAM}{random access memory}
\newacronym{bram}{BRAM}{block RAM}
\newacronym{dvfs}{DVFS}{dynamic voltage frequency scaling}
\newacronym{rtl}{RTL}{register transfer level}
\newacronym{gwp}{GWP}{global warming potential}
\newacronym{hpc}{HPC}{high performance computing}
\newacronym{pg}{PG}{power gating}
\newacronym{cg}{CG}{clock gating}
\newacronym{ltm}{LTM}{leveraging thermal margin}
\newacronym{vpr}{VPR}{versatile place and route}
\newacronym{sb}{SB}{switch box}
\newacronym{ac}{AC}{approximate computing}
\newacronym{ast}{AST}{abstract syntax tree}
\newacronym{dsp}{DSP}{digital signal processing}
\newacronym{nic}{NIC}{network interface card}
\newacronym{tco}{TCO}{total cost of ownership}
\newacronym{avm}{AVM}{accelerated virtual machine}
\newacronym{vm}{VM}{virtual machine}
\newacronym{rcsp}{RCSP}{resource-constrained scheduling problem}
\newacronym{alu}{ALU}{arithmetic logic unit}
\newacronym{qos}{QoS}{quality-of-service}
\newacronym{crac}{CRAC}{computer room air conditioning}

\title{A Survey of FPGA Optimization Methods\\ for \rev{Data Center Energy Efficiency}}

\author{Mattia~Tibaldi and Christian~Pilato,~\IEEEmembership{Senior~Member,~IEEE}
\thanks{M. Tibaldi and C. Pilato are with the Dipartimento di Elettronica, Informazione e Bioingegneria, Politecnico di Milano, Milan, Italy (Contact email: \textit{mattia.tibaldi@polimi.it}).}}

\begin{document}

\maketitle
\begin{abstract}
This article provides a survey of academic literature about \gls{fpga} and their utilization for energy efficiency acceleration in data centers. The goal is to critically present \rev{the existing \glspl{fpga} energy optimization techniques and discuss how they can be applied to such systems. To do so,} the article explores current energy trends and their projection to the future with particular attention to the requirements set out by the \textit{European Code of Conduct for Data Center Energy Efficiency}. The article then proposes a complete analysis of over ten years of research in energy optimization techniques, classifying them by purpose, method of application, and impacts on the sources of consumption. Finally, we conclude with the challenges and possible innovations we expect for this sector.
\end{abstract}


\section{Introduction}
\label{sec:introduction}
The total amount of data generated globally is rapidly increasing and expecting to reach 180 zettabytes by 2025~\cite{statista}. This trend is due to several factors. For example, in 2020, the amount of data created and replicated reached a new high due to the \textit{COVID-19} pandemic and the boost of smart-working and contact tracing~\cite{covid}. \rev{Nowadays,} this massive amount of data is processed almost entirely in large data centers, generating up to 2\% of the global $CO_2$ emissions~\cite{9291049,9291205}. Since one of the key challenges for the next years is certainly climate change, the \textit{\gls{it}} sector must necessarily contribute to reducing emissions. In this context, it is no longer possible to ignore the data center's contribution, and analyzing the impact of modern technologies is extremely important to make educated decisions for their sustainable development~\cite{Rene20}.

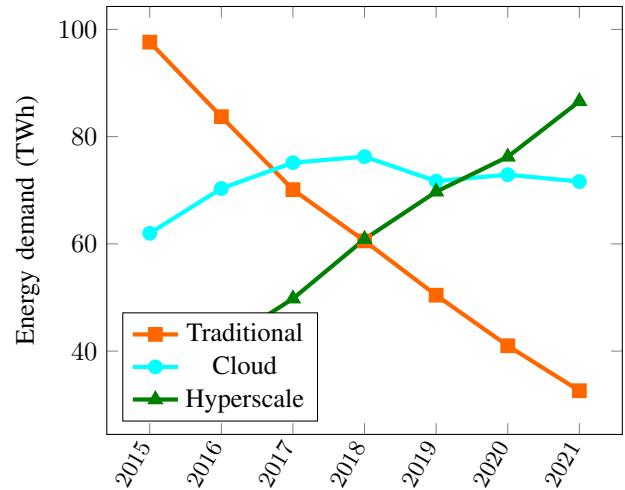
\begin{figure}
\centering
\begin{tikzpicture}
    \begin{axis}[
        legend pos=south west,
        xtick=data,
        xticklabel style=
        {/pgf/number format/1000 sep=,rotate=60,anchor=east,font=\small},
        ylabel=Energy demand (TWh)]
    \addplot[color=my_orange,mark=square*, ultra thick] coordinates {
        (2015,97.62)
        (2016,83.72)
        (2017,70.11)
        (2018,60.55)
        (2019,50.42)
        (2020,41.00)
        (2021,32.61)
    };
    \addplot[color=my_cyan,mark=otimes*, ultra thick] coordinates {
        (2015,61.97)
        (2016,70.33)
        (2017,75.14)
        (2018,76.27)
        (2019,71.70)
        (2020,72.90)
        (2021,71.62)
    };
    \addplot[color=my_green,mark=triangle*, ultra thick] coordinates {
        (2015,31.11)
        (2016,41.21)
        (2017,49.78)
        (2018,60.87)
        (2019,69.72)
        (2020,76.23)
        (2021,86.58)
    };
    \addlegendentry{Traditional}
    \addlegendentry{Cloud}
    \addlegendentry{Hyperscale}
    \end{axis}
\end{tikzpicture}
\caption{Data Center power demands from 2015 to 2021 \cite{statista22}}
\label{fig:datacenterPower}
\end{figure}

A data center is a physical structure where hundreds to thousands of servers are allocated, organized, and managed to provide specific no-stop services. Data centers can be classified by the type of architecture, i.e., \textit{traditional}, \textit{cloud}, and \textit{hyperscale}, and by the \textit{tier} level, i.e., from 1 to 4, based on their characteristics, like uptime guarantee, downtime per year, redundancy, concurrently maintainable, and price~\cite{6946229}. 
A \textit{traditional} data center is a small set of \gls{it} equipment and on-premises, often in conjunction with a corporate office. A \textit{cloud} data center combines physical servers running a single operating system with virtualized ones. In this way, a single physical server can house multiple virtual servers, increasing the efficiency and scalability of the infrastructure. When this concept reaches the limit, we have \textit{hyperscale} data centers. Hyperscale systems can manage a large network of computers (e.g., tens of thousands) to automatically respond to changes in user requests and assign workloads on demand to computing, storage, and network resources. Figure~\ref{fig:datacenterPower} shows the consumption trends for each data center category. Hyperscale data centers are becoming predominant, and their overall energy demand reached 86.58~TWh in 2021~\cite{statista22}. Instead, more and more companies are abandoning traditional data centers in favor of cloud or hyper-scale versions. 
Table~\ref{tab:tier} summarizes the major characteristics of each \textit{tier}. We need to keep this hierarchy in mind when we compare it from a data center energy standpoint. Typically, the higher \textit{tiers} are also the most \rev{consuming devices}.

To reduce the energy consumption of data centers, \gls{fpga} is emerging as a computing technology in this field. With the Catapult project~\cite{Putnam14} in 2014, Microsoft introduced \glspl{fpga} in commercial systems. Many \gls{it} protagonists such as Alibaba, Amazon, and Huawei now support \glspl{fpga} in their data centers and make them available to application developers. Today a market of over 6 billion dollars is estimated for this technology~\cite{fpgaMarket22}. However, their long-term adoption in these venues is not guaranteed. Their simple integration into data centers does not mean significant reductions in energy consumption. Managing them requires specific middleware, hardware virtualization, and domain separation mechanisms that make designing efficient architectures for such systems complex~\cite{Pilato2021}. \rev{Also, it is not always true that \gls{fpga} consumes less than \gls{cpu} or other accelerators like \gls{gpu}. Only a combination of specific problems with specific complexity results in immediate energy saving~\cite{8735546}.} Several optimizations must be provided for \glspl{fpga} to \rev{make a sustainable system}. The configuration also introduces significant challenges. Traditional energy optimization techniques require function calibrations and must re-evaluate their effectiveness. Many obstacles still exist in \gls{fpga} data center deployment. The main one is the lack of support for \glspl{fpga} in the asset management and data center monitoring tools. Besides, there is not enough software stack to allow easy deployment on the cloud. This survey analyzes this problem, discussing which solutions can be applied at each level and highlighting the open points \rev{and the expected benefits}.
After describing the energy targets for the next year and the \textit{European Code of Conduct for Data Center Energy Efficiency} (Section~\ref{sec:targets}), and providing an \rev{analysis on the current status of the technology} (Section~\ref{sec:background}), we present our main contributions:
\begin{itemize}
\item An \rev{overview} of the main metrics and methods used in data centers to analyze energy consumption (Section~\ref{sec:metrics}). 
\item A \rev{critical review of} the energy optimization techniques on \glspl{fpga} and their \rev{possible} integration in data centers (Section~\ref{sec:technique}). This part is the core of this work, summarizing over ten years of research \rev{and presenting suggestions on to apply the methods and increase their effectiveness}. 
\item A discussion of the most used solutions and the possible innovations (Section~\ref{sec:powerTrends} and Section~\ref{sec:futureInnovation}, respectively). 
\end{itemize}
\rev{The entire article is structured to address and cover all the points presented in the \textit{Code of Conduct} concerning the introduction of new technologies in data centers. We indeed use this manifest to guide the reader through the work.}
\rev{Finally,} we conclude the article in Section~\ref{sec:conclusion} with a summary of the opportunities and challenges in this domain.

\begin{table}[!t]
\renewcommand{\arraystretch}{1.3}
\centering
\caption{Data Center Tiers Compared}
\label{tab:tier}
\vspace{-6pt}
\begin{tabular}{@{}m{1.8cm} | m{1.3cm} m{1.3cm} m{1.3cm} m{1.3cm} @{}}
\toprule
 \multicolumn{1}{l}{\textbf{Parameters}} &
 \multicolumn{1}{l}{\textbf{Tier 1}} &
 \multicolumn{1}{l}{\textbf{Tier 2}} &
 \multicolumn{1}{l}{\textbf{Tier 3}} &
 \multicolumn{1}{l}{\textbf{Tier 4}} \\
\midrule 
 Uptime guarantee & 99.671\% & 99.741\% &  99.982\% & 99.995\%\\
 Downtime per year & $<$ 28,8 h  & $<$ 22 h & $<$ 1,6 h & $<$ 26,3 min\\
 Redundancy & None & \rev{Partially} & \rev{Partially+} & Fully\\
 Concurrently maintainable & No & No & Partially & Fully\\
 Price & \$  & \$\$ & \$\$\$ & \$\$\$\$\\
 \hline
 Typical customer & Small companies & Medium business & Large businesses & Government entities\\
\bottomrule
\end{tabular} 
\end{table}

\section{Energy targets for the next years}
\label{sec:targets}
\begin{figure}
\centering
\includegraphics[width=0.65\columnwidth]{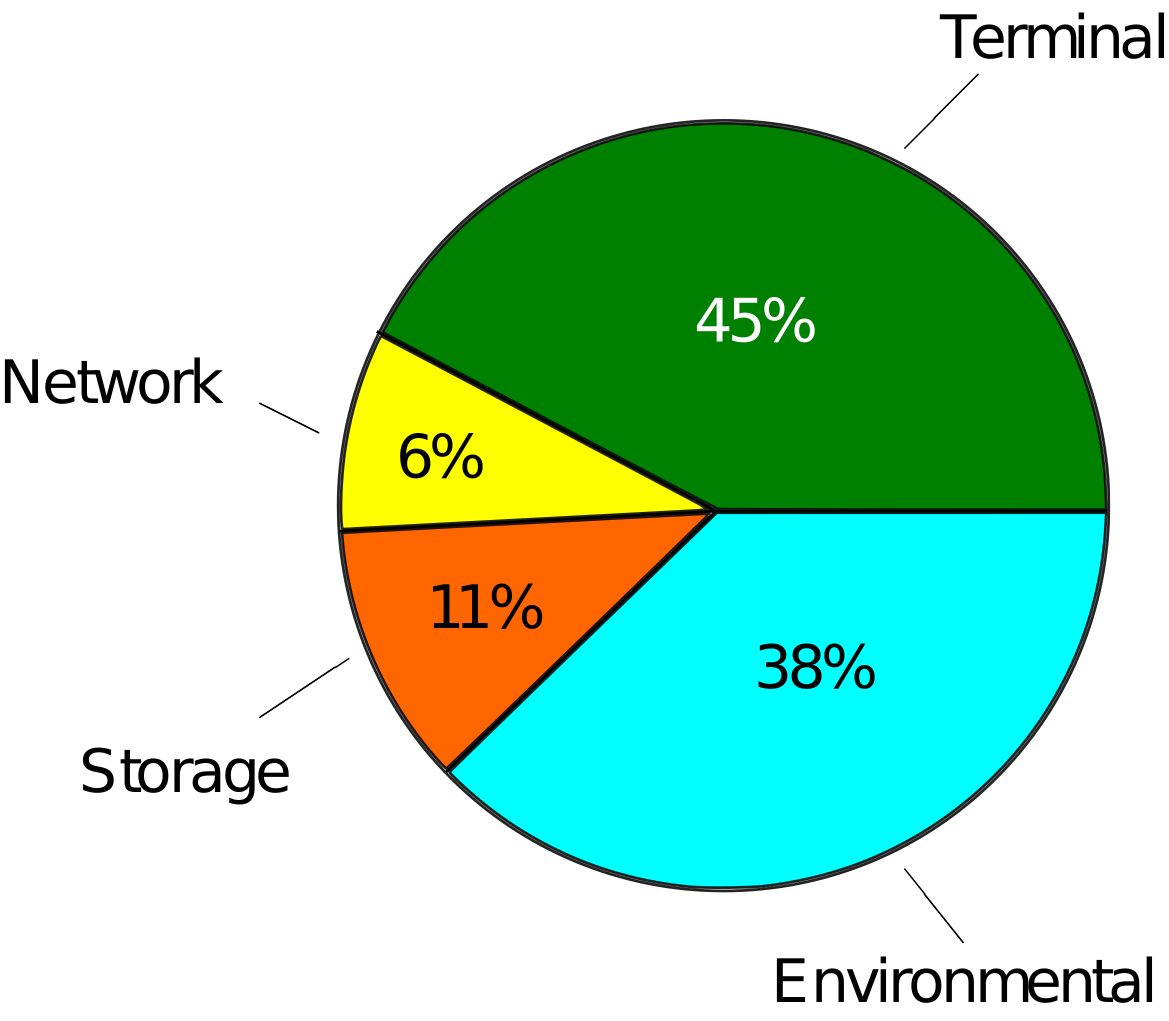}
\caption{Main sources of consumption in a data center.}
\label{fig:powerCategory}
\end{figure}

  Works like~\cite{Honee12,9265961,9028165} identify and classify the biggest causes of energy consumption in data centers. The energy consumption devices can be generally classified into four categories:
\begin{itemize}
    \item \textbf{Terminal devices}: e.g., \glspl{cpu}, \glspl{gpu}, \glspl{fpga}, and servers;
    \item \textbf{Network devices}: e.g., routers and switches;
    \item \textbf{Storage devices}: e.g., memories, \glspl{ssd}, and \glspl{hdd};
    \item \textbf{Environmental devices}: e.g., cooling, lighting, and power supply devices;
\end{itemize}
Figure~\ref{fig:powerCategory} shows how these categories impact data center consumption. We attribute most of the consumption to the terminal and environmental devices, representing 45\% and 38\% of the total absorbed energy, respectively~\cite{7279063}. The ideal condition is that all the absorbed energy is used entirely by the \gls{it} devices. Currently, few data centers manage to get close to this ideal condition, and they succeed only thanks to their particular \rev{geographical} position. For example, the \textit{Norwegian Green Mountain} data center~\cite{greenMountain} exploits the geological conformation of the environment and the harsh climate of the fjords as a cooling system, almost eliminating the energy consumption due to environmental devices. Unfortunately, not all data centers can benefit from such advantageous locations and other methods are needed to limit their consumption.

In 2008, \rev{a voluntary initiative, named \textbf{European Code of Conduct for Data Center Energy Efficiency program}~\cite{Avgerinou17}, was created within the European Union} in response to the increasing energy consumption in data centers and the need to reduce the related environmental, economic, and energy supply security impacts. Over 120 organizations participate in the program, and 289 data centers periodically submit complete energy data to the initiative. The \textit{European Code of Conduct for Data Center Energy Efficiency} poses a series of guidelines and best practices to help ensure that participants are committed to a substantial energy-saving effort. They associate each solution to an application area (e.g., entire data center, new software, new \gls{it} equipment, new build or retrofit, and optional) and a value from 1 to 5 to indicate the level of benefit to be expected and the relative priorities. In the following, we report some of the guidelines for \gls{it} equipment reported with the highest priority:
{\it
\begin{enumerate}[leftmargin=5.5mm]
    \item[\rev{\ding{202}}] Include the energy-efficiency performance of the \gls{it} device as a high-priority decision factor in the tender process;
    \item[\rev{\ding{203}}] Include the operating temperature and humidity ranges at the air intake of new equipment as high-priority decision factors in the tender process;
    \item[\rev{\ding{204}}] Formally change the deployment process to include the enabling of power management features on \gls{it} hardware as it is deployed;
    \item[\rev{\ding{205}}] Select equipment that provides mechanisms to allow the external control of its energy use;
    \item[\rev{\ding{206}}] Processes should be put in place to require senior business approval for any new service that requires dedicated hardware and will not run on a resource-sharing platform;
\end{enumerate}
}
\noindent Note that new software is reported as \textit{fundamental} to make the energy use of the software a primary selection factor.

\Gls{fpga} technology fully meets the requirements for new \gls{it} devices by the code of conduct, becoming a potential investment for most of the data centers in the area. However, their simple integration does not guarantee optimal results in reducing consumption. Specific knowledge is needed on the main existing optimization techniques to fully exploit them, \rev{and professionals from different disciplines must work together to ensure satisfactory results. On this aspect, the \textit{Code of Conduct} focuses on the importance of establishing an \textquote{approval board} for each decision including representatives from all disciplines like senior management, IT, engineering, applications/software, and procurement. Collaboration is essential to properly understand the problem and reach an effective solution in such complex systems. This survey aims to be a reference tool for deciding whether to invest in \glspl{fpga} within data centers and is structured to respond to each caveat presented in the \textit{Code of Conduct}. Section \ref{sec:background} highlights the current state of the \gls{fpga} technology with the latest innovations. Section \ref{sec:metrics} discusses potential comparison metrics between different solutions giving hints for defining new ones. Each of the five points above (i.e., \ding{202} to \ding{206}) is reflected in Section~\ref{sec:technique}, where we evaluate existing energy optimization techniques. For completeness, we have preferred to discuss all the existing energy optimizations on FPGAs and give an opinion on their adoption in the data center rather than limiting ourselves only to the most promising techniques. Each optimization follows two classifications: one based on the expected benefits in terms of energy savings (Table~\ref{tab:effort}) and one based on which area of the data center it affects (Figure~\ref{fig:allPowertechniques}). } 

\section{\gls{fpga} \rev{innovations and active researches}}
\label{sec:background}

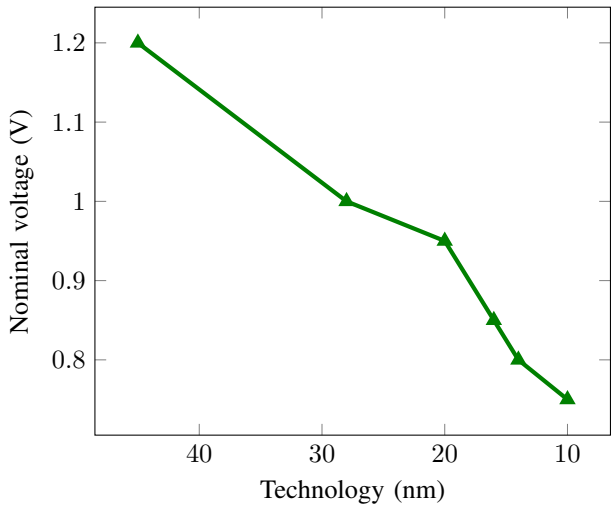
\begin{figure}
\centering
\begin{tikzpicture}
    \begin{axis}[
        xlabel=Technology (nm),
        ylabel=Nominal voltage (V),
        x dir=reverse]
    \addplot[color=my_green,mark=triangle*, ultra thick] coordinates {
        (45,1.20)
        (28,1.00)
        (20,0.95)
        (16,0.85)
        (14,0.80)
        (10,0.75)
    };
    \end{axis}
\end{tikzpicture}
\caption{Voltage trends for the main production processes.}
\label{fig:voltageTrend}
\end{figure}

\gls{fpga} devices are reconfigurable chips based on CMOS and \gls{sram} technologies. \rev{Their structure achieves excellent performance and low consumption. On these devices, one can create a customized implementation of an algorithm that operates on multiple data in parallel. So, it can be faster and consume less power than processors with higher clock speeds. Reconfiguration allows users to change the functionality every time is needed. So, an FPGA trade offs performance and flexibility, resulting a suitable product for data centers where the same node can serve multiple users. In this section, we highlight the main shortcomings of the technology with today's solutions and possible research areas to improve their use inside data centers.}

\rev{\subsection{The need for efficiency}}
\rev{The integration level of FPGA technologies is currently far from the \gls{cpu} world, resulting in reduced  efficiency. Great progress has been made since the first \gls{fpga} from Xilinx appeared on the market in 1985. This initial technology was with a production process at 2000 nm and the first devices had 64 \gls{clb}. Today, after almost 40 years of research, we find \glspl{fpga} on the market with over 100,000 \glspl{clb} made with production processes ranging from 45~nm for AMD-Xilinx Spartan 6~\cite{xilinxSpartan} to 10 nm for Intel-Altera Agilex~\cite{alteraAgilex}. Having a clear understanding of the type of technology and its evolution is essential to identify the optimizations and challenges of the future, especially in terms of energy. Also, the \textit{Code of Conduct} calls for considering the entire hardware manufacturing process to understand and estimate which impacts the technology will have inside the infrastructure. Different production processes correspond to different supply voltages and different consumption. \Cref{fig:voltageTrend} shows the linear trend of the voltages for the main production processes, from 1.2~V for 45~nm down to 0.75~V for 10~nm. Improving the production process allows designers to reduce the consumption of these devices. }

Recently, the media announced the Intel/VMware Crossroads 3D-FPGA Academic Research Center as a multi-university effort to improve the future of \gls{fpga} technology~\cite{IntelFPGAacc}. By stacking multiple \gls{fpga} dies vertically, researchers should be able to achieve a higher transistor density while also balancing performance, power, and manufacturing costs. Therefore, the research in this area is far from standing, and there are ample opportunities for improvement.

\rev{\subsection{The need for bandwidth}} 
\rev{Due to the proliferation of big data applications~\cite{Pilato2021}, the need to move greater quantities of data has resulted in increasing bus lines inside the \gls{fpga}.} One of the main problems that \glspl{fpga} suffered was the latency for memory access and the poor throughput they obtained. With the recent introduction of \textbf{\glspl{hbm}}, the problem has been partially addressed~\cite{Jun2017}. \gls{hbm} is a 3D-stacked \gls{dram} that offers high-bandwidth and energy-efficient data movements introduced in the latest generation \glspl{fpga} such as AMD Xilinx's Alveo U280 and Intel Altera's Stratix 10. This type of memory allows reaching throughput of over 460 GB/s, in the case of the Alveo U280, allowing fast memory access and reducing the effect of \gls{sll}, \rev{the high latency technology currently used to access FPGA resources from any of its regions~\cite{Choi2020}. However, the HBM technology with large bus lines is difficult to be fully exploited, demanding hard-crafted solutions or advanced design methods~\cite{trets2021,iris2023}.}

\rev{\subsection{The need for improved clock structures}
Conventional FPGA core architectures have been based on balanced clock trees, which minimize deterministic skew. This method has served well for designs up to 500 MHz, but they need innovative solutions to reach speeds of up to 1 GHz. The solution must minimize local variation and skew, and provide a flexible network to serve the numerous clock regions. For example, to address these challenges, Intel introduces an entirely new core architecture, called \textit{Intel HyperFlex FPGA Architecture}~\cite{HyperFlex}. They add additional registers in every interconnect routing segment and at the inputs of all functional blocks. With these elements, they can retime registers to eliminate critical paths, add pipeline registers to remove routing delays, and optimize the design for best-in-class performance.}

\rev{\subsection{The need for simplified development}}
Creating an \textquote{implementation} for \gls{fpga} consists in making a circuit that relates to the resources present in the device. The process uses \gls{hdl} such as Verilog or VHDL or, more commonly today, C/C++ language compiled with an \gls{hls} compiler for the circuit definition. We can identify four steps for generating the \gls{fpga} configuration file (i.e., \textit{bitstream}).
\begin{enumerate}
    \item \textbf{Hardware design}: \gls{hdl} codes are manually written manually or derived from high-level specifications by means of \gls{hls} compilers;
    \item \textbf{Synthesis}: \gls{hdl} codes are compiled and translated into netlists;
    \item \textbf{Implementation and routing}: the synthesized design is mapped onto the target \gls{fpga} resources and the connections are defined among them;
    \item \textbf{Bitstream generation}: the implemented design is translated into a configuration file (\rev{also called \textquote{bitstream}}) which can be downloaded onto the \gls{fpga}.
\end{enumerate}
Using \gls{hdl} languages is not simple, and realizing well-optimized projects requires substantial specific knowledge and time. Major \gls{fpga} manufacturers provide development kits that simplify this tedious and error-prone process. Both \textit{AMD-Xilinx Vitis suite} and \textit{Intel-Altera Quartus suite} contain \gls{hls} compilers that allows the designer to transform a program written with a high-level language like C or C++ into \gls{hdl}. These tools allow designers to apply several optimizations such as pipelining and loop unrolling, making application development easier and faster~\cite{7368920} by means of specific \textit{pragmas} (e.g., \textbf{\#pragma hls dataflow}, \textbf{\#pragma hls array partition}, or \textbf{\#pragma hls inline}) that are used by the designers to annotate the code in order to specify where to apply such optimizations. In addition to these solutions, there are many open-source \gls{hls} projects such as Bambu~\cite{Ferrandi2021}, Merlin~\cite{CongHPW016}, and AutoDSE \cite{Sohrabizadeh22} that try to make this process even more intuitive and automated for a programmer without specific knowledge about \glspl{fpga}. Recently, there has been a growing interest in open-source compilers, not only in academia. For example, Xilinx recently released the source code of the \textit{Vitis HLS Front-End}~\cite{vitisHLSfront} that can help boost the innovation around HLS tools. Conversely, the technical specifications of the physical devices (including the bitstream format) are still mostly closed-source. We argue that more flexibility from commercial solutions (e.g., clear API for interaction and analysis of implemented designs) can further boost research and innovation in this area.

\begin{figure*}
\centering
\begin{subfigure}{.3\textwidth}
  \centering
  \includegraphics[width=1.1\linewidth]{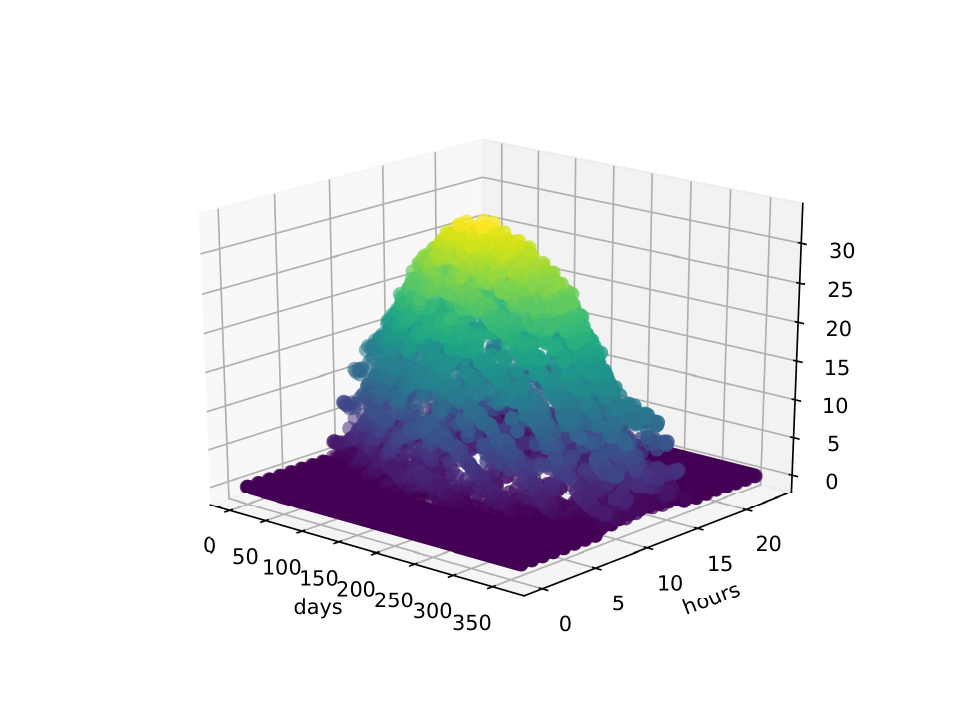}
  \caption{\rev{Solar energy production}}
  \label{subfig:sub1}
\end{subfigure}%
\begin{subfigure}{.3\textwidth}
  \centering
  \includegraphics[width=1.1\linewidth]{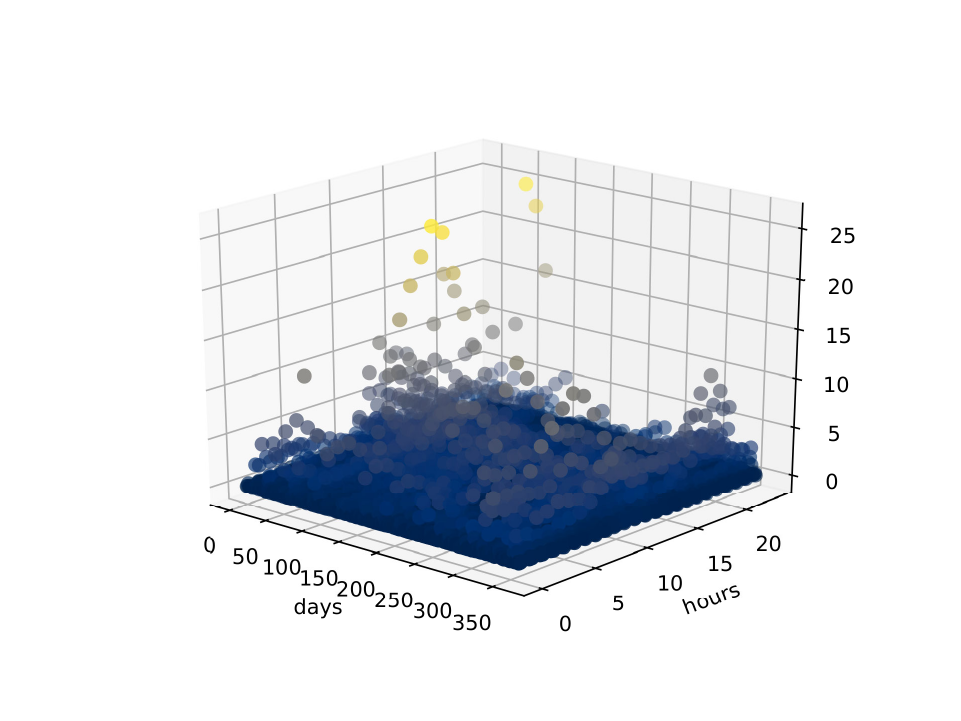}
  \caption{\rev{Wind energy production}}
  \label{subfig:sub2}
\end{subfigure}%
\begin{subfigure}{.3\textwidth}
  \centering
  \includegraphics[width=1.1\linewidth]{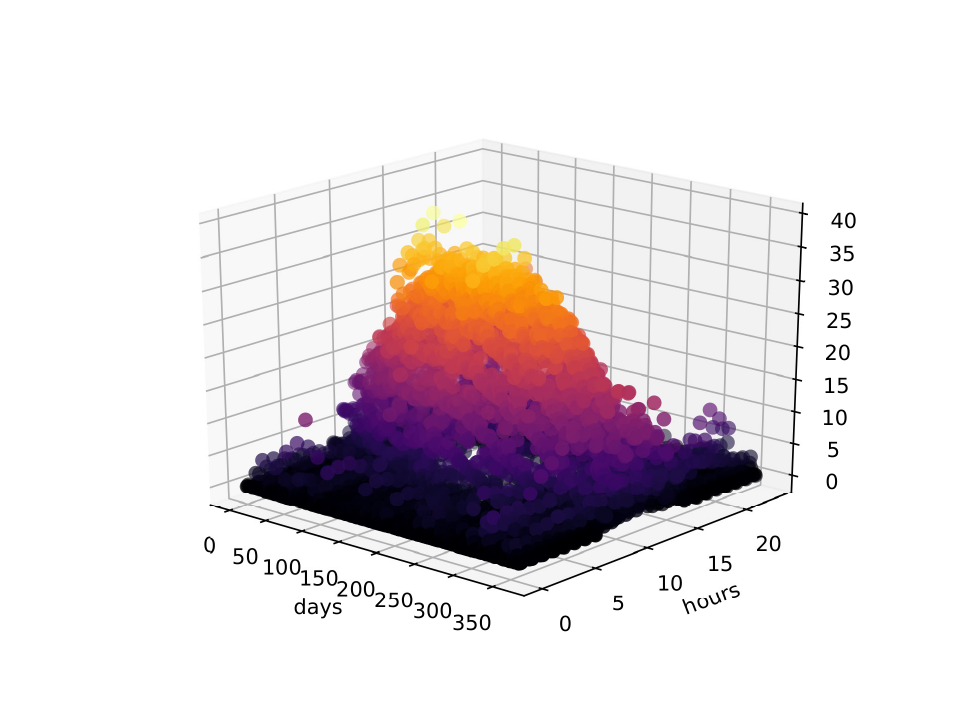}
  \caption{\rev{Combined energy production}}
  \label{subfig:sub3}
\end{subfigure}
\caption{\rev{Energy production of an hypothetical medium energy park in one year. The values are expressed in MWh.}}
\label{fig:powerProduction}
\end{figure*}

\section{Evolution of power metrics}
\label{sec:metrics}
In this section, we discuss the metrics for monitoring energy consumption on the individual \gls{fpga} device and the overall data center environment. \rev{Before starting, we want to present the distinction between power and energy consumption. Measures of power are most interesting for sizing the data center, cooling systems, or power units. On the other hand, when we talk about energy, we consider the integral of power over time. The second measurement is functional from a green point of view because it can be quickly transformed into kg of CO\textsubscript{2} produced. The \textit{Code of Conduct} suggests using both units to have complete visibility into the data center.}

\subsection{\gls{fpga} power metrics}
\Gls{cmos} transistors are the basic blocks of \glspl{fpga}; therefore, we can divide the power consumption into two categories: \textit{dynamic power} and \textit{leakage power}. \textbf{Dynamic power} occurs from the switching activities because of the short-circuiting current and charging and discharging of load capacitance. It is therefore related to when the circuit is running. As anticipated in the previous section, each \gls{clb} is logic that consumes power since the \gls{pmu} continuously powered it. \textbf{Leakage power} is always present, and the power dissipation occurs in the form of leakage current when the system is not powered or is in standby mode. The total consumption of the device is given by the sum of these two components, as shown in Equation~\ref{equ:dynamicLeakage}.

\begin{equation}\label{equ:dynamicLeakage}
    P_{fpga} = \alpha CFV^2+gV^3
\end{equation}
\noindent where $CFV^2$ represents dynamic power, and $gV^3$ is the leakage one. In addition, $\alpha$ is the activity factor, $C$ the capacitance, $V$ the supply voltage, $F$ the frequency, and $g$ the leakage factor, which is an intrinsic factor of the device under analysis. The dynamic power has a quadratic trend with increasing voltage, while the leakage power has a cubic trend. Today, modern packaging techniques allow for high densities of \gls{cmos} per unit area, making leakage power the main factor in the equation. Further considerations are covered in Section~\ref{subsec:technologyLevel}. Observing Equation~\ref{equ:dynamicLeakage}, we can adjust only two parameters to improve energy consumption. The first is the frequency which affects only the dynamic power, and the second is the voltage which instead reduces both components. In Section~\ref{sec:technique}, we discuss the techniques that operate on these two parameters, how \glspl{fpga} can integrate them, and what benefits they entail. 

Often the results of the energy consumption of the accelerators are related to the number of floating-point operations performed in the unit of time (FLOPS), leading to the definition of the FLOPS/Watt~\cite{9362524} metric. This metric allows us to effectively compare different accelerators and choose the one with a higher value (i.e., the most energy-efficient one).

The model discussed above does not take into account the thermal state of the device. A chip subjected to diverse temperatures consumes a different amount of energy, and therefore, running the same software will absorb more (less) if the chip is hotter (cooled). \textbf{Thermodynamics computing}~\cite{arXiv:1911.01968} tries to unify the mathematical model of the information with the thermodynamic model. The adoption and development of this model would lead to a better understanding of the phenomenon illustrated and the possibility of having energy models that are more accurate and better \rev{optimize} than the current ones. 
The model would also be relevant from the \textbf{design of the cooling system} viewpoint~\cite{ZHANG2021102253}. We can model the capacity of the cooling system (C) as in Equation~\ref{equ:FanModel}.

\begin{equation}
\label{equ:FanModel}
    C=\sqrt[3]{\frac{P_{fan}}{T_{hot} - T_{cold}}}
\end{equation}
\noindent Where $P_{fan}$ is the maximum power absorbed by the board fans, $T_{hot}$ is the temperature of the exhausting air from the chip, and $T_{cold}$ is the operational target temperature, commonly set to 25 degrees. Studies also show that for every watt of power utilized during the chip operation, the cooling equipment consumes an additional 0.5-1 watts to extract the exhausting air from the \gls{it} racks~\cite{7279063}. This example shows how desirable a model change is and how it fits into the design of the entire data center. In conclusion, optimizations that can optimize absorbed energy and dissipated heat simultaneously are the most suitable for future developments.

\subsection{Data center power metrics}\label{subsec:dcPowerMetric}
As mentioned in section~\ref{sec:targets}, the causes of energy consumption in a data center are many, and we can individuate four categories: terminal devices, network devices, storage devices, and environmental devices. The sum of all these components gives the overall consumption of the data center. \cite{Avgerinou17} reports the average energy consumption of 289 European data centers participating in the code of conduct, estimating an average annual electricity consumption of 13,684 MWh, of which 7,871 MWh refer only to the \gls{it} sector (e.g., terminal devices and network devices). In a data center, monitoring systems, included within the power distribution units (PDUs), continuously control consumption. Each rack has one or more dedicated PDUs with several outputs equal to the number of servers allocated (e.g., up to 42 servers per rack). In real time, it is possible to know how much energy is absorbed by a specific server, rack, or cluster (e.g., a set of racks). This solution identifies potential faults in the system and allows the technician to keep consumption under control. Consumption data are collected and processed by specific tools that monitor the progress of the infrastructure. Google is working on new tools that link consumption with environmental impact and give complete transparency to the public on what happens in a data center to raise public awareness of the issue~\cite{tools21}.

In this work, we focus on the consumption due to the \gls{it} department and, in particular, the terminal devices. The consumption of a server can be modeled as the consumption of the individual parts \cite{8807458}, combined as shown in Equation~\ref{equ:serverPower1}.

\begin{align}
\label{equ:serverPower1}
    P_{server} = &P_{cpu}u_{cpu}+P_{memory}u_{memory}+P_{disk}u_{disk}+\\
    \notag &P_{nic}u_{nic}+P_{fpga}u_{fpga}
\end{align}

Where $P_{cpu}, P_{memory}, P_{disk}, P_{nic}$, and $P_{fpga}$ are the power consumption of the respective components, while $u_{cpu}, u_{memory}, u_{disk}, u_{nic}$, and $u_{fpga}$ are the utilization rate of the different components. In the ideal case, the utilization of the accelerators should be one, so it is always exploited to the maximum, reducing the utilization of more power-hungry components, such as the \gls{cpu}.

A different model, instead, treats the data center similar to the one for \gls{fpga}~\cite{8807458}, dividing the power into dynamic $P_{var}$ and static $P_{fix}$. We can model it as follows:

\begin{equation}
\label{equ:DCPower}
    P_{total} = P_{fix} + P_{var}
\end{equation}
\noindent The idle states of the servers characterize $P_{fix}$, while the moments of use and the actions that occur, like the phases of data computation or movement between physical memories, characterize $P_{var}$.

The \textit{European Code of Conduct for Data Center Energy Efficiency} defines, in addition to the total power consumption in Watts, a second metric called \textbf{\gls{pue}}~\cite{Avgerinou17}, introduced by The Green Grid in 2007. The \gls{pue} helps understand the data center's efficiency and how to reduce energy consumption. It is the ratio between the total data center input power and the power used by the \gls{it} equipment (e.g., Equation~\ref{equ:pue}).
\begin{equation}
\label{equ:pue}
    PUE = \frac{Total Facility Power}{IT Equipment Power}
\end{equation}
The ideal value is 1, which happens when the entire facility power is consumed by the \gls{it} department and not by lighting, cooling systems, \glspl{pdu}, fans, and other equipment. Higher values imply lower efficiency.  The average \gls{pue} of today's European data centers is 1.8, but the geographic location of the individual facility is important. For example, thanks to its particular structure and position, the \textit{Green Mountain} data center~\cite{greenMountain} has a \gls{pue} of 1.2.

\rev{Concerning point \ding{202}, the \textit{Code of Conduct} suggests the introduction of new metrics that distinguish between the energy consumed from renewable and fossil sources. Renewable energy is not continuous in time and can assume decidedly different values according to the period. For example, consider a solar panel. It will produce more during the summer and sunny hours, rather than during the winter and nights. Figure~\ref{fig:powerProduction} shows this variability in the two principal renewable sources exploited today by studying the energy production of a hypothetical wind and solar farm on an annual and daily scale. Figure \ref{subfig:sub3} shows the total renewable energy entering the data center as a combination of the two production sites. An interesting metric could consider these production curves and estimate the over-consumption (i.e., how much non-renewable energy the data center uses) and the under-consumption (i.e., how much renewable energy the data center is produced but not used). Equation \ref{equ:greeenMet} defines this metric:}

\begin{equation}
\label{equ:greeenMet}
    \rev{M(E_u) = \begin{cases} \alpha\frac{E_r - E_u}{E_r}, & \mbox{if } E_u \le E_r \\ \beta\frac{E_u - E_r}{E_u}, & \mbox{if } E_u > E_r \end{cases}}
\end{equation}
\rev{Where $E_u$ is the instant used energy, $E_r$ is the instant renewable energy, and $\alpha | \beta$ are two parameters between 0 and 1 to scale the relevance of the contribution. The data center better uses the available energy when $M$ is close to 1. This metric can be combined with adaptive systems that regulate accelerator energy consumption (see Section~\ref{subsec:appLogicLevel}).}

The metrics that we have presented do not consider the consumption due to the construction of the data center and its components, as well as do not relate energy consumption with environmental metrics such as the \gls{gwp}~\cite{Fthenakis06}~\cite{Pacca02}~\cite{Unece21}. A model that considers all these aspects is the \textbf{\gls{lca}}. \gls{lca} is a powerful tool applicable to the data center sector~\cite{Shah12}. In latest years, some works have been carried out on the subject. \cite{Whitehead15} present a complete \gls{lca} from cradle to grave of a UK data center. Their work highlights how the impact on the environment of a data center is completely attributable to production and the use of \gls{it} equipment (mainly servers). They also observe how the gap between the impacts of production and use decreases in the states where energy production has greener. \cite{Honee12} proposed a similar study on a Swedish data center, concluding that new emerging technology can improve the environmental performance of the \gls{it} equipment. Currently, in the literature, no work proposes an \gls{lca} on \glspl{fpga}, reducing only to emphasizing their energy benefits.

\section{Power Optimization Techniques}
\label{sec:technique}
\begin{figure*}
\centering
\includegraphics[width=\textwidth]{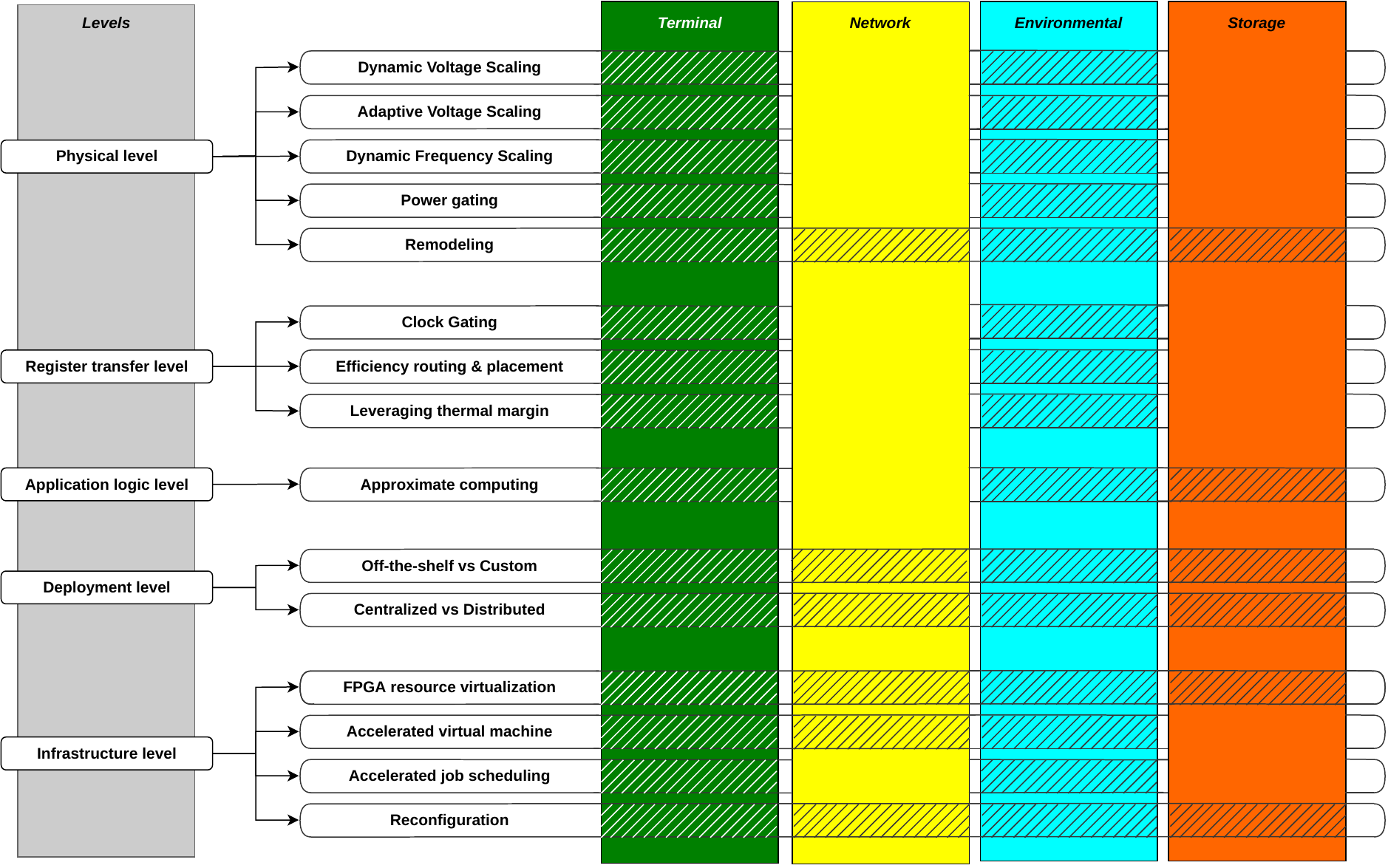}
\caption{Summary of the energy optimization techniques described in section \ref{sec:technique}. For each technique, the categories of consumption on which they affect, both negative or positive influence, have been highlighted through the lined background of the boxes.}
\label{fig:allPowertechniques}
\end{figure*}

\rev{While there are several techniques to regulate the energy consumption of \glspl{fpga}, not all of them are suitable or attractive in data centers}. In this section, we present the fundamental solutions classified by levels, from the lowest \textit{physical level} to the highest \textit{infrastructure level}\rev{, analyzing their relevance for the data center}. For each subsection, we present a summary table of the techniques described with the substantial works that implement them and a complete discussion on how the optimization impacts the overall data center infrastructure and power consumption. Figure~\ref{fig:allPowertechniques} summarizes the energy optimization techniques described in the section, highlighting the categories of energy consumption in which they affect\rev{, while Table~\ref{tab:effort} classifies the techniques by importance.  In particular, we assign \ding{93}\ding{93}\ding{93} to techniques that guarantee substantial energy savings and are directly connected to the \textit{Code of Conduct}, \ding{93}\ding{93} to complementary solutions implementable in data centers with moderate energy savings, and \ding{93} to optional techniques.}

\begin{table}[]
\caption{\rev{Expected benefits on the entire data center by applying each energy optimization. \ding{93}\ding{93}\ding{93} identifies techniques that guarantee substantial energy savings and are directly connected to the \textit{Code of Conduct}; \ding{93}\ding{93} identifies complementary solutions for data centers with moderate energy savings, while \ding{93} identifies optional techniques.}}
\label{tab:effort}
\center
\begin{tabular}{m{10cm} m{0.2cm} m{0.2cm} m{0.2cm} m{0.2cm}}
\multicolumn{5}{c}{\textbf{\rev{Physical level}}}                                         \\ \hline
\multicolumn{2}{r|}{\textit{\rev{Dynamic voltage scaling}}}        & \multicolumn{3}{l}{\rev{\ding{93}}} \\
\multicolumn{2}{r|}{\textit{\rev{Adaptive voltage scaling}}}       & \multicolumn{3}{l}{\rev{\ding{93}}} \\
\multicolumn{2}{r|}{\textit{\rev{Dynamic frequency scaling}}}      & \multicolumn{3}{l}{\rev{\ding{93}}} \\
\multicolumn{2}{r|}{\textit{\rev{Power gating}}}                   & \multicolumn{3}{l}{\rev{\ding{93}\ding{93}}} \\
\multicolumn{2}{r|}{\textit{\rev{Remodeling}}}                     & \multicolumn{3}{l}{\rev{\ding{93}\ding{93}\ding{93}}} \\
\multicolumn{1}{l}{}                      &                  &       &       &      \\
\multicolumn{5}{c}{\textbf{\rev{Register transfer level}}}                                \\ \hline
\multicolumn{2}{r|}{\textit{\rev{Clock gating}}}                   & \multicolumn{3}{l}{\rev{\ding{93}\ding{93}}} \\
\multicolumn{2}{r|}{\textit{\rev{Efficient routing \& placement}}} & \multicolumn{3}{l}{\rev{\ding{93}}}\\
\multicolumn{2}{r|}{\textit{\rev{Leveraging thermal margin}}}      & \multicolumn{3}{l}{\rev{\ding{93}\ding{93}}} \\
\multicolumn{1}{l}{}                      &                  &       &       &      \\
\multicolumn{5}{c}{\textbf{\rev{Application logic level}}}                                \\ \hline
\multicolumn{2}{r|}{\textit{\rev{Approximate computing}}}          & \multicolumn{3}{l}{\rev{\ding{93}\ding{93}}} \\
\multicolumn{1}{l}{}                      &                  &       &       &      \\
\multicolumn{5}{c}{\textbf{\rev{Deployment level}}}                                       \\ \hline
\multicolumn{2}{r|}{\textit{\rev{Off-the-shelf vs Custom}}}        & \multicolumn{3}{l}{\rev{\ding{93}\ding{93}\ding{93}}} \\
\multicolumn{2}{r|}{\textit{\rev{Centralized vs Distributed}}}     & \multicolumn{3}{l}{\rev{\ding{93}\ding{93}\ding{93}}} \\
\multicolumn{1}{l}{}                      &                  &       &       &      \\
\multicolumn{5}{c}{\textbf{\rev{Infrastructure level}}}                                   \\ \hline
\multicolumn{2}{r|}{\textit{\rev{FPGA resource virtualization}}}   & \multicolumn{3}{l}{\rev{\ding{93}\ding{93}\ding{93}}} \\
\multicolumn{2}{r|}{\textit{\rev{Accelerated virtual machine}}}    & \multicolumn{3}{l}{\rev{\ding{93}\ding{93}}} \\
\multicolumn{2}{r|}{\textit{\rev{Accelerated job scheduling}}}     & \multicolumn{3}{l}{\rev{\ding{93}\ding{93}}} \\
\multicolumn{2}{r|}{\textit{\rev{Reconfiguration}}}                & \multicolumn{3}{l}{\rev{\ding{93}\ding{93}}}
\end{tabular}
\end{table}

\subsection{Physical level}
\label{subsec:technologyLevel}

In \Cref{sec:metrics}, we discussed the main metrics to measure the energy consumption of \glspl{fpga}, defining the dynamic and the leakage power. For chips made with production processes below 90 nm, it turns out that the leakage component prevails over the dynamic one. The smaller the process is, the more the leakage component is overcome~\cite{Karel14}. For today's products, optimizations that work on leakage power will be more effective than those that operate on dynamic power. Unfortunately, the techniques that can adapt to reduce the leakage power are few, and the most advantages are applicable only at a low level since the leakage power is inseparable from the technology used. At the physical level, we find the following energy optimization techniques\rev{, all of them are directly linked to point \ding{202}}: 

\begin{itemize}
    \item \textit{\Gls{dvs}} allows to manipulate the voltage with which the \gls{fpga} is powered. Requires careful calibration at design time;
    \item \textit{\Gls{avs}} is based on the same idea as \gls{dvs} but the calibration takes place at run time with a monitoring system;
    \item \textit{\Gls{dfs}} allows to adjust the frequency and set it to its optimal value;
    \item \textit{Power gating} turns off logic when not in use;
    \item \textit{Remodeling} redesigns how the \glspl{fpga} are internally structured with an optical push for energy optimization;
\end{itemize}

\noindent \textbf{Dynamic voltage scaling} involves reducing the supply voltage of a circuit. It can reduce both dynamic and leakage current, but at the expense of increasing circuit delay, which can lead to timing violations and, in turn, errors~\cite{6823413}~\cite{4380689}. This effect happens because the voltage acts as the "accelerator pedal" for the signals to propagate. If we have a low voltage and a high clock frequency, parts of the circuit may work with the wrong values because of the delay. For best results, the final design should operate at the voltage that reduces power consumption as much as possible while maintaining the circuit working. 
Finding this threshold is problematic since the optimum operating voltage changes with time and between devices. For this reason, the system must be carefully calibrated \rev{to not} run into problems. \gls{dvs} is a best practice in \gls{asic} design where the circuit does not change, and there are no further complications (e.g., timing violation due to the modification of the electric schema and a wrong \gls{dvs} configuration). In \glspl{fpga}, on the other hand, more attention is needed since re-configuring the device also requires a \gls{dvs} re-calibration to avoid timing problems with the risk of having an unreliable circuit. 
Chow et al. and Qi et al. ~\cite{1568543}~\cite{7929183} propose a methodology for supporting \gls{dvs} on commercial \glspl{fpga}. Their idea estimates the delay of critical paths by implementing a \gls{ldmc} inside the \gls{fpga}. The module, through feedback signals, interfaces directly with the \gls{pmu} to regulate the \gls{fpga} voltages. This solution effectively reduces energy consumption by up to 54\% with negligible resource consumption. One advantage of their methodology is that it does not require additional design effort or changes to the \gls{fpga} itself. The main limitation of this approach is that it requires experimentation to find appropriate threshold values for each \gls{fpga}, and it can markedly lengthen development time. Ahmed et al.~\cite{7577342} reduce the time due to the calibration of the \gls{ldmc} by developing an automated tool (FRoC). FRoC ensures that the calibration process is invisible to \gls{fpga} users and does not add any extra manual steps to the design process. The tool generates a calibration table to scale the voltage while the application is active. A similar approach is also demonstrated by the authors of Nunez-Yanez et al.~\cite{4380689}. Unlike the others, they apply the reverse procedure. First, they reduce voltage and then search for the correct operating frequency. Significant savings in power and energy are measured from the nominal value for the Virtex-4 Xilinx \gls{fpga} of 1.2 V down to its limit of 0.9 V.

A different approach to the problem is given by \textbf{Adaptive Voltage Scaling}. In this case, it eliminates the calibration problem by introducing a monitor to measure the performance variability of the silicon. Works such as those proposed by Nabina and Nunez-Yanez~\cite{2392618}~\cite{6945345} show how these systems operating in a closed-loop configuration can significantly improve energy profiles compared with \gls{dvs}. Implementing this control system directly within the \gls{fpga} can limit the available resources. To get around the problem, Nunez-Yanez implements the control and monitoring system on an external PCB connected to the \gls{fpga} board. The results are good, and it will be useful if \gls{fpga} vendors move towards making their devices more friendly to \gls{avs} techniques by adopting different power planes and introducing more robust built-in delay monitoring circuits. The works we have just presented do not consider applications that use the \gls{fpga} hard blocks such as \glspl{bram}. Zhao et al.~\cite{8533470} extend a previously proposed offline calibration-based \gls{dvs} approach to enable \gls{dvs} for \glspl{fpga} with \glspl{bram}. Again the idea is to run a series of tests to ensure that all used \gls{bram} cells operate safely while scaling the supply voltage.

An optimization that always goes hand in hand with \gls{dvs} is \textbf{dynamic frequency scaling}, leading to the combined acronym \gls{dvfs}. As we have seen, when we applied \gls{dvs}, it can run into timing problems due to the high frequencies that the circuit should work. Adding \gls{dfs} serves to solve this problem, thus, being able to optimize voltage and frequency together. 
\gls{dfs} can adjust the frequency and set it to its optimal value. The addition of \gls{dfs} to \gls{dvs} results in additional linear power savings~\cite{4252714}. Zhao et al.~\cite{7468125} presents an offline self-calibration scheme, which automatically finds the \gls{fpga} frequency and core voltage operating limit. Their idea is to explore two phases, one to estimate the actual minimum frequencies at which the logic can operate and one to generate the configuration tables for the \gls{ldmc} module. These phases require one reprogramming of the \gls{fpga} each, and a power saving of around 40\% justified the additional time spent in the design phase. 

No doubt, the most effective technique for reducing leakage power on a chip is \textbf{power gating}. The idea is simple, turn off logic when not in use. Often, the literature uses the term \textit{dark silicon} to refer to areas of the chip switched off~\cite{7927201}. The power gating technique inserts a power switch unit, controlled by an on/off signal, on the logic to optimize. The power switch's inclusion introduces overhead on the resources and the energy. The fundamental challenge for any power gating technique is to ensure that the saved leakage power outweighs the power overhead of the power gating. 
We can identify two different implementation styles, the \textit{fine-grain} (divides the circuit into small sections) and the \textit{coarse-grain} (divides the circuit into large sections) technique. 
The two styles are usually combined to seek a trade-off between resources used and energy saved~\cite{5681533}. The more finely implemented the control system is, the more the overhead will be relevant because it is necessary to insert more power switches in the circuit. Finding the right compromise can be difficult. In \glspl{fpga}, we can safely turn off all the unused resources by a specific design without any risk. It is always possible to perform this optimization with negligible overhead since for how the placement and routing phases on \gls{fpga} are made, the logic consumed is allocated in a single contiguous region, outlining a clear separation between used and unused sections.
Therefore, by inserting a single power switch, you can turn off all the unused areas of the \gls{fpga} by the implemented design~\cite{4114899}. 

There are particular designs where the \textit{coarse-grain} style has a greater area overhead than \textit{fine-grain}. This overhead happens when the logic zones we want to control are very far from each other, resulting in long paths that add overhead in the signal distribution network~\cite{5483137}. On the other hand, in \textit{fine-grain} power gating, the higher level of control allows to reduce the leakage power more. Still, since the power switches are always running, they increase the consumption of the dynamic one. Due to these overheads, \textit{fine-grain} power gating is less efficient than coarse-grain power gating~\cite{4114899}~\cite{6404219}~\cite{4068935}.

Another challenge with power gating is to reactivate the switched-off logic without creating delay when needed. This consideration is fundamental when we work with stream applications where the stream must proceed without interruption. A possible solution is to add a series of checkpoints along the computation phases of the streams to reactivate the logic in time~\cite{7293946}. 
\gls{dvs}, \gls{avs}, and power gating are techniques not always applicable to commercial \glspl{fpga}. To implement these techniques, the \gls{pmu} must be accessible and configurable from an external component such as a \gls{cpu} or an \gls{asic} associated with the \gls{fpga}. The level of control we have over the \gls{fpga} and how we can implement these techniques depends on the \gls{pmu} structure. Many of the papers presented propose to \textbf{remodel} \glspl{fpga} to implement energetic optimization techniques more effectively. Qi et al.~\cite{8308753} redesign the \glspl{fpga} internal structure with an optical push for energy optimization.  
They offer a custom \gls{fpga} that includes native solutions of \gls{dvfs} and Power \rev{g}ating, eliminating many of the previously discussed issues (e.g., interconnect delay, calibration, and so on). Today, we are shifting towards custom \gls{fpga} versions\rev{, also in relation to point \ding{204} and \ding{205}. Few other techniques consider these aspects and these new architectures will be taken into consideration more and more}. This increase can be seen both in the embedded systems field and in the \gls{hpc} field. Projects such as \textit{OpenFPGA}~\cite{9098028} facilitate these new architectures prototyping. \textit{OpenFPGA} allows users to customize their \gls{fpga} architectures down to circuit-level details using a high-level architecture description language and to autogenerate associated Verilog netlists which can be used in a backend flow to generate production-ready layouts. We expect this trend to increase, and in the future, we will probably have \gls{fpga} models with specific targets for the use sectors. Table~\ref{tab:technologySummary} summarise the optimization techniques and the main works presented in this subsection.

\vspace{4pt}
{\sc OVERALL IMPACTS CONSIDERATION: }\textit{The optimization techniques presented in this subsection have the main purpose of limiting the consumption due to leakage power, acting on the voltage of the \gls{fpga} and the working frequency of the circuit. The primary effect obtained is reducing the $P_{fpga}$ and, in turn, the energy consumed by the device terminals. The impacts of these techniques on the environment are not easy to deduce without a unified model with thermodynamics that considers \glspl{fpga}. Based on the Equation~\ref{equ:FanModel} and the approximation for \gls{crac} cooling systems which says that for every 1 Watt saved on the chip, we also save about 1 Watt from the cooling system~\cite{ZHANG2021102253}, we can still conclude that there is an impact and that it is positive. \rev{However, even though these techniques impact both \ding{202} and \ding{203}, we believe that \gls{dvs}, \gls{avs}, and \gls{dfs} are valuable solutions for embedded systems but not for data centers. Therefore, we prefer to consider them optional. On the other hand, power gating is different, which we think is a good practice and which the new generation synthesis tools are adopting by default.} As far as the remodeling technique is concerned, a broader discussion is needed since it could also impact the network and storage component based on how the re-modeling is carried out. A remodel pushed towards network management by inserting specific resources for the network inside the \gls{fpga} could have a significant impact on the network devices. In contrast, a remodel that leads the \glspl{fpga} to have more types and storage capacities could favorably impact the second component. Currently, there are no studies relating the remodel to the consumption categories. Still, we \rev{ consider it an extremely good technique that touches on every aspect listed in the code of conduct.}}

\begin{table}[!t]
\renewcommand{\arraystretch}{1.3}
\centering
\caption{Physical level - Power optimization techniques} 
\label{tab:technologySummary}
\vspace{-6pt}
\begin{tabular}{@{}m{2.5cm} m{2.5cm} c c @{}}
\toprule
 \multicolumn{1}{c}{\textbf{Techniques}} &
 \multicolumn{1}{c}{\textbf{Main works}} &
 \multicolumn{2}{c}{\textbf{Target power}} \\
   \cline{3-4}
& & \textbf{Dynamic} & \textbf{Leakage} \\
\midrule 
 \rev{Dynamic voltage scaling} & \cite{7577342} \cite{1568543} \cite{6823413} \cite{7110584} \cite{4380689} \cite{7929183} & \ding{51} & \ding{51}\\
 \rev{Adaptive voltage scaling} &  \cite{8533470} \cite{2392618} \cite{6945345} & \ding{51} & \ding{51}\\
 \rev{Dynamic frequency scaling} &  \cite{7110584} \cite{4252714} \cite{7468125} & \ding{51} & \ding{51}\\
 Power gating &  \cite{5681533} \cite{7293946} \cite{5483137} \cite{4114899} \cite{7927201} \cite{6404219} \cite{4068935} & \ding{55} & \ding{51}\\
 Remodeling &  \cite{8308753} \cite{9098028} & \ding{51} & \ding{51}\\
\bottomrule
\end{tabular} 
\end{table}

\subsection{Register Transfer Level}
\label{subsec:rtl}
\Gls{rtl} is an abstraction level for defining the digital portions of a design. It is the principal abstraction used for specifying electronic systems today. \Gls{hdl} like Verilog and VHDL uses \gls{rtl} to create high-level representations of a circuit, from which specific tools can derive lower-level abstraction, now integrated into the most common development kits such as Vivado and Quartus. 
Thanks to this level of abstraction, it is possible to conduct an early-stage dynamic power analysis~\cite{569539}. The optimization techniques available at this level focus on reducing the dynamic power \rev{(point \ding{202})}, and they operate without having to access specific components external to the \gls{fpga}, as happens instead for the physical layer that needs the \gls{pmu}.
The dynamic power is due to the switching activity of the signals, introducing a continuous charge and discharge of the parasitic capacitance present in the circuit~\cite{7529097}.
Typically the signal with the highest switching activity is the clock signal. This sentence is especially true when we consider synchronous digital systems. At the register transfer level, we find the following energy optimization techniques: 

\begin{itemize}
    \item \Gls{cg}: disables the clocking of specific registers when the outputs of those registers are stable; 
    \item Energy-efficient routing and placement: improve the interconnections between \gls{fpga} resources to guarantee energy savings;
    \item \Gls{ltm}: tries to exploit the \gls{fpga} temperature to drop further the voltage; 
\end{itemize}

\noindent There is one principal technique that acts on the switching activities, and it is \textbf{clock gating}. As the term suggests, clock gating is a way to reduce switching activity on circuit signals by disabling the clocking of specific registers when the outputs of those registers are stable~\cite{cta.2107}~\cite{5272538}. The idea of the clock-gating technique for \glspl{asic} has developed in the late nineties. Studies indicate that the clock signals in digital computers consume 15-45\% of the system power~\cite{841927}. Because of its effectiveness, clock-gating has been a hot topic in many research areas~\cite{7860113}~\cite{6211782}~\cite{1690091}. Modern \glspl{fpga} include several clock control blocks that allow to shutting down of the clock line in some parts of the circuit, and to emulate the functionality of the gated clocks is possible to use feedback multiplexers~\cite{4664863}. Pandey et al.~\cite{6533362} and Liong Tan et al.~\cite{8703530} implement an energy-efficient \gls{alu} on \gls{fpga}. Both authors agree on the impossibility of adopting the traditional \gls{asic} style to implement \gls{cg} with AND gate because it will create a glitch in output. In modern \gls{fpga} devices, two buffer types can replace the AND gate and ensure glitch-less output: Global Clock Buffer (BUFGCE) and HROW Clock Buffer (BUFHCE). Building \gls{cg} using these buffers results in an overhead that includes extra control logic to generate \gls{cg} control signal and extra leakage power consumption. However, the technique saves 40\% of the dynamic power~\cite{6658025}. The same idea seen for clock gating can also be applied to other design-dependent signals and even to the memory structure, as shown by the works of Sterpone et al.~\cite{5763128} and Agrawal et al.~\cite{7878231}. Geier et al.~\cite{9470464} present a module that can easily be added to current designs with memory-mapped AXI3 or AXI4Stream (AXI4S) interfaces to monitor interface signals and limit switching activity. The module insertion is completely transparent to the design and provides a quick method of applying clock gating on \glspl{fpga}.

Specific tools such as the one presented by Zhang et al.~\cite{6406915} or the one developed by Siemens~\cite{powerPro} can even perform automatically \rev{c}lock gating optimization. For example, \textit{powerPro}~\cite{powerPro}, starting from an \gls{rtl} file and a set of vector tests runs a series of simulations to verify in the design when and where it is possible to disable the clock signal. The execution times depend on the design size and can range from a few minutes to several hours, but they are acceptable given the results. Research into automating these optimizations is critical today. With increasingly stringent power consumption requirements, they help to provide solutions already in the early stages of implementation.

Another technique that always acts on the circuit is \textbf{energy-efficient routing and placement}. This technique improves the interconnections between \gls{fpga} resources to guarantee energy savings~\cite{799437}~\cite{8064897}. Essentially, the closer the resources are to each other, the less energy is consumed. Exploiting some design rules like avoiding the use of \gls{sll} or fast tracks and using local interconnection can save power dissipation in  \glspl{fpga}~\cite{1488567}. Zemani and Esmaili~\cite{1226332} try to do just that by extending the \gls{vpr} routing algorithm~\cite{Betz99} with one more iteration where power is optimized together with area and circuit performance. Adding different performance metrics such as those reported in Section~\ref{sec:metrics} can extend the exploration algorithm. Something similar is also done by the Hao Hoo team~\cite{6645548} and Leming and Nepal~\cite{5236058}, with the only difference being that they propose a new version for the embedded \gls{sb} of the \gls{fpga}. Seifoori et al.~\cite{8977905} are the first to use a machine learning technique to design power gating regions in the \gls{fpga} routing network. They define similarity metric, cluster pattern, and power gating efficiency to design three clustering algorithms based on K-means clustering. Finally, they evaluate the obtained design on an Intel Stratix-IV \gls{fpga}, obtaining 1.4$\times$ higher savings to other heuristics. Generally, an exploration algorithm based on the power consumption model combined with the place and route tool can help to find the optimal solution~\cite{Betz99}. Chtourou et al.~\cite{6841956} instead of focusing on routing algorithms analyze two different routing architectures: the \textit{SB\_tristate} and the \textit{SB\_multiplexer}. The first uses bidirectional \gls{sb} implemented with back-to-back tri-state drivers. The second uses bidirectional \gls{sb} implemented using tri-states and multiplexers. They conclude that \textit{SB\_multiplexer} has a significant impact on power saving compared to \textit{SB\_tristate}. This difference happens because \textit{SB\_tristate} always uses the switch in only one direction, thus increasing the amount of the leakage of power dissipated by routing resources.

The last technique explained at the register transfer level is \textbf{leveraging thermal margin}. Compared to the other optimizations presented, \gls{ltm} is much more recent and is a cross between physical layer techniques such as \gls{dvs} and energy-efficient routing and placement~\cite{Amouri13}~\cite{Zhao18}. The most recent work is  Khaleghi et al.~\cite{8988683} in 2019. The basic idea is that a chip at different temperatures exhibits different voltage limits, and they use this gap to optimize the voltage. If the chip is "cold" (e.g., 40 degrees), it is possible to drop further in voltage compared to one that is "hot" (e.g., 100 degrees). It is necessary to organize the logic on the \gls{fpga} and thus improve the existing routing and placement algorithms to avoid excessive temperature peaks. The technique in its current state is still complicated and requires several simulations (e.g., delay simulation, \gls{fpga} architecture simulation, and thermal simulation) before arriving at a stable solution. Furthermore, if you want to use an adaptive approach at run time, it requires access to temperature sensors that are not always available. However, the technique is promising, improving \gls{dvs} up to 30\%.
Table~\ref{tab:rtlSummary} summarise the optimization techniques and the main works presented in this subsection.

\vspace{4pt}
{\sc OVERALL IMPACTS CONSIDERATION: }\textit{Considerations similar to those made previously on the impact on terminal devices and environmental devices are also valid for the techniques set out in this subsection. The work done by Khaleghi et al.~\cite{8988683} allows us to expand the discussion on the \gls{fpga}'s impact on environmental devices. Looking at the thermal resistance $\theta_{JA}$, we note that for today's Intel and Xilinx \gls{fpga} products, a value $\theta_{JA}$ of $2^\circ C/Watt$ and a pessimistic thermal resistance of $12^\circ C/Watt$ are considered. If now we combine these considerations with equation \ref{equ:FanModel}, we can model $T_ {hot}$ as $P_ {fpga} \times \theta_{JA}$ and obtain a consistent model to estimate the cooling capacity required starting from a simple energy consumption profile. \rev{This consideration acquires importance, given points \ding{203} and \ding{205}, by providing innovative metrics and relevant considerations for the dimensioning of cooling systems and data centers in general. For this reason, we consider \gls{ltm} a technique that can influence more the data center than efficient routing \& placement but close to clock gating since tools like~\cite{6406915} already adopt it as a standard. Maybe in the future, thermal optimizations will be more extensively implemented into synthesis tools.}}

\begin{table}[!t]
\renewcommand{\arraystretch}{1.25}
\centering
\caption{RTL - Power optimization techniques summary}
\label{tab:rtlSummary}
\vspace{-6pt}
\begin{tabular}{@{}m{2.5cm}  m{2.5cm} c c @{}}
\toprule
 \multicolumn{1}{c}{\textbf{Techniques}} &
 \multicolumn{1}{c}{\textbf{Main works}} &
 \multicolumn{2}{c}{\textbf{Target power}} \\
  \cline{3-4}
 & & \textbf{Dynamic} & \textbf{Leakage} \\
\midrule 
 Clock gating & \cite{9470464} \cite{7878231} \cite{7860113} \cite{cta.2107} \cite{5272538} \cite{6211782} \cite{4664863} \cite{6658025} \cite{6533362} \cite{5763128} \cite{8703530} \cite{841927} \cite{1690091} & \ding{51} & \ding{55}\\
 \rev{Efficient routing \& placement} & \cite{6841956}  \cite{5236058} \cite{8977905} \cite{8064897} \cite{Betz99} \cite{1226332} & \ding{51} & \ding{51}\\
 \rev{Leveraging thermal margin} & \cite{Amouri13} \cite{8988683} \cite{Zhao18} \cite{6645548}& \ding{51} & \ding{51}\\
\bottomrule
\end{tabular} 
\end{table}

\subsection{Application Logic Level}
\label{subsec:appLogicLevel}
By "Application Logic Level" optimizations, we refer to all the techniques that modify the application algorithm or the input/output interfaces to achieve energy efficiency \rev{(point \ding{202})}. In general, code optimizations at a high level have a greater impact on performance than low-level optimizations~\cite{4101125}. A modification to the algorithm can generate a very different circuit and exhibit a different energy profile. These techniques go by the name of \textbf{\gls{ac}}. Approximate computing is based on the intuitive observation that while performing an exact computation requires a high amount of resources, granting selective approximation can provide extreme gains in efficiency~\cite{Mittal:2016:STA:2891449.2893356}. Many of today's applications that we find in data centers (e.g., machine learning applications~\cite{6506085}~\cite{8964974}, signal processing applications~\cite{7847611}, data analytics applications~\cite{dataAda}, and so forth) can effectively manage a certain degree of approximation without running into completely wrong computations. For example, for a k-means clustering algorithm, allowing a classification accuracy loss of 5\% can save up to 50x energy~\cite{6754190}. 
Without going into too much detail, the main approximation techniques that we find in the literature are:
\begin{itemize}
    \item \textit{Custom data format}: creates smaller components by changing the precision (bit-width) of input or intermediate data to save both leakage and dynamic power~\cite{Sampson:2011:EAD:1993498.1993518};
    \item \textit{Custom operator}: generates approximate adders and multipliers to perform partial computation with effects similar to those obtainable with \textit{custom data format}~\cite{7926980}~\cite{Li15dac}~\cite{7850168};
    \item \textit{Loop perforation}: skips some iterations of a loop to reduce computational overhead and so dynamic power~\cite{Sidiroglou-Douskos:2011:MPV:2025113.2025133};
    \item \textit{Memoization}: stores the results of functions for later reuse to skip the portion of code in the second run~\cite{6617694}. This technique reduces the dynamic power but increases the leakage one. It requires some balance to be applied with confidence;
    \item \textit{Task skipping}: skips memory references, tasks, or input portions to achieve energy efficiency~\cite{Goiri15}. Similar idea and effects as \textit{loop perforation};
    \item \textit{Data sampling}: samples data from the input queue to speed up the execution~\cite{Tiba22}. Similar idea and effects as \textit{loop perforation};
    \item \textit{Custom memory hierarchy}: adds more memory layers to hide the cache miss latency~\cite{Sparsh14}. It reduces the leakage power due to data transmission over long distances, but by introducing new logic, it still increases the overall consumption. Also, in this case, the technique must be carefully balanced to obtain advantages;
    \item \textit{Multiple inexact program version}: utilizes multiple versions of application code with different trade-offs between accuracy and overheads (e.g., execution time, power consumption) and selects at run time which one is better to use~\cite{Tiba22};
\end{itemize}
Applying \gls{ac} is not always easy and requires wisely choosing the portion of code where to intervene and the technique not to have an unacceptable loss of quality. Further, careful monitoring of the output is required to ensure that quality specifications are met. New metrics are needed to estimate the error we are introducing when we use multiple techniques together (e.g., Monte Carlo simulation~\cite{Su18dac}). The work presented by Nepal et al.~\cite{6800575} combines \textit{application logic level} and \gls{rtl} techniques in a framework called ABACUS. ABACUS starts by creating an \gls{ast} from the \gls{rtl} description, and after it applies AC functions (e.g., custom data type, custom operator, and loop perforation) to create fair approximate designs. At last, it identifies the most suitable design along the Pareto frontier that represents the trade-off between accuracy and power consumption by exploring the space with all the possible variants. Only variants with a specific \gls{qos} are considered acceptable. Developing frameworks that lower the difficulty and the knowledge necessary to optimize power consumption is fundamental for the sector's evolution. Other research groups such as Chandrasekaran and Amira~\cite{4101125}, Segal et al.~\cite{6927442}, and Gao et al.~\cite{Gao17iccad}  presented frameworks capable of implementing approximation techniques and estimating their effectiveness from the point of view of performance and energy consumption. Specifically, Gao and Qu suggest a runtime framework to exploit runtime energy information. The basic idea is to use a low-cost method like the error-resilient characteristics of each operator to estimate the impact of immediate input values on the accuracy of computation and then decide whether directly use the approximated value or perform an accurate computation. In general, \gls{ac} is a powerful technique that allows saving from 40\% up to 80\% of the power consumption. However, if the data to be processed and the accelerator does not share the same data format, most of these benefits disappear, and some conversion is necessary~\cite{8745716}. In our opinion, conversion between data formats is an expensive operation from both an energy and time point of view; and it represents one of the biggest challenges within data centers where the vast heterogeneity of data often requires conversions. Table~\ref{tab:allSummary} summarise the optimization techniques and the main works presented in this subsection.

\vspace{4pt}
{\sc OVERALL IMPACTS CONSIDERATION: }\textit{Approximate computing reduces energy consumption by approximating the logic behind the processing algorithms. This approach impacts storage, terminal devices, and environmental ones. Many approximation techniques affect data and how it is read from memory. We remind you that the consumption of the memories depends on the amount of memory installed and the number of reads and write accesses made in the unit of time. Custom data format, memoization, and custom memory hierarchy are examples of optimizations that change the amount of memory used. Custom data format reduces memory usage by simplifying the data format to be more compact in memory. The technique is, therefore, also effective in reducing consumption by the storage component. On the contrary, memoization and custom memory hierarchy increase the amount of memory required and negatively affect the storage component of the data center. Optimizations such as data sampling, on the other hand, act on the number of accesses in the memory, guaranteeing higher energy savings, similar to the case just discussed with the custom data format. \rev{In conclusion, we consider the technique advantageous and applicable to data centers, especially when the main target is reinforcement learning or deep learning applications, where these techniques are almost mandatory. As mentioned in Section \ref{sec:metrics}, multiple inexact program versions can be combined with innovative metrics to monitor and intervene quickly and autonomously on the accelerator, keeping its efficiency under control.}}

\begin{table}[!t]
\renewcommand{\arraystretch}{1.25}
\centering
\caption{Application logic Level - Power optimization techniques summary}
\label{tab:allSummary}
\vspace{-6pt}
\begin{tabular}{@{}m{2.9cm} m{2.5cm} c c @{}}
\toprule
 \multicolumn{1}{c}{\textbf{Techniques}} &
 \multicolumn{1}{c}{\textbf{Main works}} &
 \multicolumn{2}{c}{\textbf{Target power}} \\
 \cline{3-4}
  & & \textbf{Dynamic} & \textbf{Leakage} \\
\midrule 
 Approximate computing &  \cite{4101125} \cite{Gao17iccad} \cite{8745716} \cite{6506085} \cite{7926980} \cite{Li15dac} \cite{Tiba22} \cite{Mittal:2016:STA:2891449.2893356} \cite{6800575}  \cite{7847611} \cite{6927442} \cite{7850168} & \ding{51} & \ding{55}\\
\bottomrule
\end{tabular} 
\end{table}

\subsection{Deployment level}\label{subsec:deployment}
This subsection discusses the dominant data center boards currently on the market and how data centers installed them. \rev{It represents an excellent guide to the choice of the product and its installation in the data center, covering fundamental aspects given by the \textit{Code of Conduct}, such as points \ding{204}, \ding{205}, and \ding{206}.} A data center is normally structured in areas called clusters. Racks are the base component of each cluster, and they can house up to 42 servers. Each rack is powered by a redundant \gls{pdu} placed at the bottom of the structure and by at least one network switch for accessing the primary network. A secondary network often exists alongside the primary network, which gives direct access to other racks in the cluster or between the clusters themselves. Servers integrate \glspl{fpga} with PCI-e connection.

\begin{figure}
\centering
\begin{tikzpicture}
  \begin{axis}[
    xbar, xmin=0,
    width=6.5cm, height=8cm, enlarge y limits=0.1,
    xlabel={mWatt per DSP},
    symbolic y coords={
    FALCON Stratix 10,
    PAC Arria 10GX,
    PAC N3000,
    Alveo U200,
    Alveo U280,
    Alveo U30,
    IPU C5020X,
    PAC D5005,
    Alveo U250,
    Alveo U55C,
    Alveo U50
    },
    ytick=data,
    nodes near coords, nodes near coords align={horizontal},
    ]
    \addplot[fill=my_cyan] coordinates  {
    (48.5,FALCON Stratix 10)
    (39.1,PAC Arria 10GX)
    (32.9,PAC N3000)
    (32.9,Alveo U200)
    (24.9,Alveo U280)
    (21.7,Alveo U30)
    (18.9,IPU C5020X)
    (18.6,PAC D5005)
    (18.3,Alveo U250)
    (16.6,Alveo U55C)
    (12.6,Alveo U50)
    };
  \end{axis}
\end{tikzpicture}
\caption{Power consumption trends of the main off-the-shelf \gls{fpga} Data center boards.
}
\label{fig:boardTrend}
\end{figure}
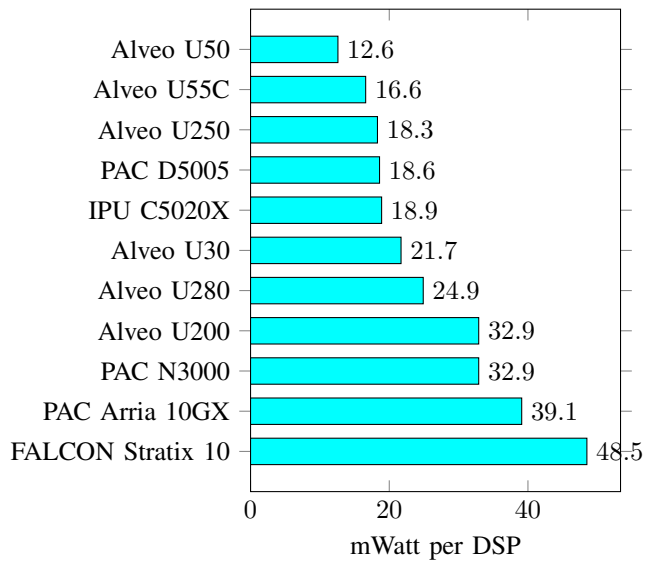

Currently, no cloud provider creates its \glspl{fpga}. Hence, the smallest unit of differentiation is the \gls{fpga} board both \textbf{Off-the-shelf} and \textbf{Custom} are possible~\cite{3506713}. The possible economic advantage depends on the quantities and the type of use. Table~\ref{tab:boardSpecification} collects information on eleven off-the-shelf products between AMD Xilinx and Intel Altera. Each one is a general-purpose board with specific features (e.g., \gls{dsp}, \gls{ram}, and \gls{hbm}). The technologies with access to \glspl{hbm} are the most promising for the \gls{hpc} field since they have a high throughput at low power consumption. In recent years, Xilinx has created the product line Alveo specific for the data center environment. However, we do not find such a clear distinction for Altera products where the Stratix 10 is probably the best solution. Altera currently produces its \glspl{fpga} with a more advanced lithographic technology than Xilinx, although on paper, Xilinx appears to perform better in terms of power consumption with a ratio of 0.025~Watt/DSP for the Alveo U280 versus a ratio of 0.048~Watt/DSP for the FALCON Stratix 10. Figure~\ref{fig:boardTrend} shows the energy consumption normalized on the number of \glspl{dsp} embedded in the board. On the other hand, with custom \gls{fpga} boards, any feature can be varied, such as form factor, cooling, memory type and size,  and \gls{fpga} family. This customization ensures that the boards closely match the requirements of the target system, which in many cases are stringent. We rarely encounter data center platforms designed natively for \glspl{fpga}; instead, we easily find upgrades of existing architectures with specific constraints on power supply, temperatures, form factor, and cooling system. Furthermore, in the custom solutions, the \gls{pmu} is accessible from the \gls{cpu}, allowing \rev{physical level optimization and better monitoring of the accelerator that perfectly matches the point \ding{205}}. For these systems, it is difficult, if not impossible, to find such specific boards on the market, and therefore, the design of a custom board is the only solution~\cite{Putnam15}~\cite{Dieter20}. For this reason, it would be appropriate to move towards boards that can be assembled through modules, thus reducing the need to develop custom boards.

\begin{table}[!t]
\renewcommand{\arraystretch}{1.4}
\centering
\caption{Main off-the-shelf \gls{fpga} Data center boards specification}
\label{tab:boardSpecification}
\vspace{-6pt}
\begin{tabular}{@{}c c c c c@{}}
\toprule
 \multicolumn{1}{c}{\textbf{Products}} &
 \multicolumn{1}{c}{\textbf{Technology}} &
 \multicolumn{1}{c}{\textbf{DSP}} &
 \multicolumn{1}{c}{\textbf{RAM}} &
 \multicolumn{1}{c}{\textbf{HBM}} \\
\midrule 
 \textbf{AMD Xilinx} & &  &  & \\
 \hline
 Alveo U30 & 16~nm & 3456 & 8~GB & - \\
 Alveo U50 & 16~nm & 5952 & - & 8~GB \\
 Alveo U55C & 16~nm & 9024 & 16~GB & - \\
 Alveo U200 & 16~nm & 6840 & 64~GB & - \\
 Alveo U250 & 16~nm & 12288 & 64~GB & - \\
 Alveo U280 & 16~nm & 9024 & 32~GB & 8~GB \\
\midrule 
 \textbf{Intel Altera} & &  &  & \\
 \hline
 PAC Arria 10GX & 20~nm & 1687 & 8~GB & - \\
 IPU C5020X & 14~nm & 3960 & 16~GB &  - \\
 PAC N3000 & 14~nm & 3036 & 9~GB & - \\
 FALCON Stratix 10 & 14~nm & 3960 & 12~GB & 8~GB \\
 PAC D5005 & 14~nm & 11520 & 32~GB & - \\
\bottomrule
\end{tabular} 
\end{table}

Integrating \gls{fpga} technology into a platform does not automatically mean saving energy and having an advantage. It all depends on the type of interconnectivity the \gls{fpga} has with the other elements inside the data center and on the actual utilization. If the \gls{fpga} does not exploit its resources, what you get is simply a system with higher losses due to the introduction of new hardware~\cite{Xiaoyu19}. Bobda et al.~\cite{3506713} provide a complete analysis of the future of \gls{fpga} acceleration in data centers, describing in detail the currently most used placement techniques, but do not investigate the energy impacts that these choices entail. We can therefore identify two types of placement: \textbf{distributed} and \textbf{centralized}. Having a distributed \gls{fpga} placement means that nodes have their \glspl{fpga}, and other nodes can access it. This solution leads to greater utilization, more scalability, and more power reduction since the CPU workload is offloaded more effectively to an \gls{fpga} pool~\cite{Putnam14}~\cite{Francois17}. It is also possible to place \glspl{fpga} in a centralized manner. Typically we see this choice for network architectures, where the single \gls{fpga} is coupled to an \gls{asic} or a \gls{cpu} to implement a smart \gls{nic} or Smart Switch. Such a deployment needs fewer \glspl{fpga}; this typically translates to easier management, the lower \gls{tco}, and smaller average node sizes~\cite{3506713}. \rev{We strongly recommend the adoption of \glspl{fpga} in network devices. Due to their structure and composition, the implementation of network protocols on \glspl{fpga} is advantageous with very high performance. Furthermore, this allows the creation of custom network protocols with a significant impact on data analysis applications, or in general, on applications where most of the time is spent transferring data from storage to computing nodes.} 

Virtually all data centers today implement \glspl{fpga} in a distributed way because it ensures the always exploitation of the resources and that the idle moments of the \glspl{fpga} are at a minimum~\cite{Chung2018}. If a node does not use its local \gls{fpga}, another node can take possession of it and use it to speed up its workflow. Implementing a distributed infrastructure involves many challenges:
\begin{enumerate}
    \item the power consumption due to the introduction of many \glspl{fpga} must not exceed the amount of power saved by their use;
    \item the internal data center network infrastructure must be redesigned to allow access to \glspl{fpga} between nodes, increasing other sources of consumption;
    \item the management and monitoring are more complex, and there can be concurrency problems;
\end{enumerate}
Several papers ~\cite{George16}~\cite{Alan10}~\cite{Chris05}~\cite{Baxter07}~\cite{Thomas15} present examples of architecture of this type. However, in almost all systems, the network equipment necessary for the communication between \glspl{fpga} is implemented internally to the \glspl{fpga}, consuming resources. Developing and using new switches capable of supporting \glspl{fpga} would be a great help to the increasingly easy integration of \glspl{fpga}~\cite{Caulfield16}~\cite{7929186}. More recent lines of research focus on the third point, the management of this new resource within the data center~\cite{9439431}
 and facilitating its development~\cite{8514395}. The most famous platforms, such as Openstack~\cite{8711829} and Kubernetes~\cite{8968907}, which instantiate, and reconfigure data center resources in real-time, can effectively optimize energy consumption but do not have support for \glspl{fpga}. A possible new field of research could be to extend these tools by understanding what it means to optimize \glspl{fpga}, together with the other components already present. Table~\ref{tab:depSummary} summarise the optimization techniques and the main works presented in this subsection.

\vspace{4pt}
{\sc OVERALL IMPACTS CONSIDERATION: }\textit{At this level, the data center architecture is defined, and the choices made will impact each source of consumption presented. Each board has a different consumption and amount of memory as presented in Table~\ref{tab:boardSpecification} and Figure~\ref{fig:boardTrend}. In addition to the memory typology, there is a different kind of cooling system and network interconnection installed on the board. These features affect the consumption of the terminals, environmental, and storage devices. Custom solutions allow the designer to better balance these aspects by producing boards specific to the type of architecture designed. Of course, it is not always possible to opt for solutions of this type due to the development costs incurred. 
\rev{Also, it is essential to consider what types of workloads the data center will be computing to understand whether to adopt \glspl{fpga} in the computing nodes, in the network components, or both. Many applications have benefits in speeding up communications, and some of the computations, such as data filtering, can already be applied at the network level while also reducing the resulting traffic. }
The most commonly accepted choice of implementing \glspl{fpga} in a distributed way hurts consumption related to the network that must allow \glspl{fpga} to communicate with each other. The type of network implemented and the number of networks created define the overall impact. From the literature analyzed, we discover that the most common type of network is the torus and that a maximum of two networks are created~\cite{Putnam14}~\cite{Putnam15}~\cite{Francois17}. The first network, where the data center exchanges the data in and out, is called primary. The second is called secondary, which connects the individual \glspl{fpga}, allowing the sharing of resources. The increase in network consumption that we obtain is, therefore, a side effect of this solution which can be considered acceptable given the advantages of flexibility and use that it guarantees. \rev{We believe that particular attention should be paid to these decisions as they impact the entire structure of the data center and its final performance. The last observation we want to report concerns the type of \gls{fpga} adopted. Small \glspl{fpga}, in terms of resources, have faster development times. In about an hour, the hardware designer has the final configuration file, while for large \glspl{fpga}, such as the Alveo U280, times can go up to 15 hours. So negatively affects the overall development times. On the other hand, smaller \glspl{fpga} have lower performance than large ones, but a good network infrastructure and wise use of \glspl{fpga} in a distributed manner can compensate. In the end, we recommend using the latter solution when considering \gls{fpga} integration in data centers.}}

\begin{table}[!t]
\renewcommand{\arraystretch}{1.25}
\centering
\caption{Deployment Level - Power optimization techniques summary}
\label{tab:depSummary}
\vspace{-6pt}
\begin{tabular}{@{}m{2.5cm} m{2.5cm} c c @{}}
\toprule
 \multicolumn{1}{c}{\textbf{Techniques}} &
 \multicolumn{1}{c}{\textbf{Main works}} &
 \multicolumn{2}{c}{\textbf{Target power}} \\
   \cline{3-4}
  & & \textbf{Dynamic} & \textbf{Leakage} \\
\midrule 
 Off-the-shelf board & \cite{George16}  \cite{Alan10}  \cite{Chris05} \cite{Baxter07} \cite{Thomas15}& \ding{55} & \ding{51}\\
 Custom board & \cite{Putnam14} \cite{Putnam15} \cite{Dieter20} \cite{Francois17} & \ding{51} & \ding{51}\\
 Distributed placement & \cite{Putnam14} \cite{George16} \cite{Alan10} \cite{Putnam15} \cite{Chris05} \cite{Dieter20} \cite{Francois17} \cite{Baxter07} \cite{Thomas15} & \ding{55} & \ding{51}\\
\bottomrule
\end{tabular} 
\end{table}

\subsection{Infrastructure level}

The infrastructure level represents the highest level of abstraction that we have identified to which it is possible to apply energy optimizations that include the use of \glspl{fpga}. The techniques presented in this subsection apply to large data center areas such as clusters because they require a comprehensive understanding of the current workflow. The main idea at this level is to reduce the fragmentation of resources, thus increasing their utilization. This idea shrinks execution to fewer servers, and it is possible to shut down everything else to save energy. We are therefore talking about techniques such as:
\begin{itemize}
    \item \gls{fpga} resource virtualization: exposes the internal resources of the \gls{fpga} as if they were virtualized, favoring their full utilization;
    \item \Gls{avm} placement: determines where to allocate the \glspl{vm} according to the type of workflow and energy constraints;
    \item Accelerated Job scheduling: improves the utilization of cloud servers equipped with \glspl{fpga};
    \item Reconfiguration: reconfigures \glspl{fpga} at run time according to the state of consumption of the data center;
\end{itemize}

\textbf{\gls{fpga} resource virtualization} has always been an ambitious idea ever since the early days of partial reconfiguration~\cite{1568522}. Popular virtualization techniques used in the software do not apply to \glspl{fpga} since they do not execute sequential programs but implement parallel circuits~\cite{Intel06}. \rev{Guo et al. \cite{guo2022fpga}} identify \rev{several} techniques \rev{like \textit{slot-based allocation}~\cite{1568522}, \textit{\gls{fpga} overlays}~\cite{Hayden16}, and \textit{standalone resource sharing}~\cite{GONZALEZ2012247}.}
The first divides the \gls{fpga} into fixed areas that accommodate pre-defined circuits and take advantage of partial reconfiguration at run time to change the type of behavior. \rev{This architecture adds a layer of complexity to the management of \gls{fpga} resources.} Examples of this, extended to data center solution, could be~\cite{8672857}~\cite{7082811}~\cite{3287324.3287475}. Bobda et al.~\cite{3506713} collect many works on the subject in their survey, finding that the overhead on the consumption of resources by applying this technique does not exceed 30\%. \cite{Putnam15} made similar reasoning, where partial reconfiguration is presented as one of the most important features that \glspl{fpga} offers. They propose standards for the input/output interface to freely reconfigure the kernel without ceasing to read the \gls{fpga} input stream. The second technique instead exposes a simplified layer that increases the programmability and productivity of \glspl{fpga}~\cite{6239797}. With this type of approach, it is possible to treat \glspl{fpga} similarly to \glspl{cpu}, making it more convenient for traditional virtualization~\cite{7293996}. \rev{In standalone resource sharing, the hardware resources on the  \gls{fpga} board can be categorized based on programmability, identifying the types of resources present and classifying them into logic, connectivity, or memory resources~\cite{GONZALEZ2012247}. These can then be shared individually by generating three sharing models. Respectively we will have: configuration, bandwidth, and capacity sharing. Each of these models has different properties. Configuration sharing concerns sharing logic between multiple FPGAs to speed up the computation. Bandwidth sharing, on the other hand, aims to share network or memory connectivity to improve the movement of large amounts of data. Finally, capacity sharing goes to sharing the memory on the board.}

\gls{fpga} resource virtualization, therefore, try to increase the utilization of the \gls{fpga} by exposing its resources to more processes, thus avoiding having to use other accelerators and saving energy.
\rev{Of course, sharing \glspl{fpga} expose the cloud to security issues. Different types of attacks and countermeasures have been proposed over the years. They classify cloud \gls{fpga} attacks into: \textit{extraction attacks}, with the idea to extract information from observable quantities such as power consumption and frequency \cite{9409116} \cite{https://doi.org/10.48550/arxiv.2012.07242}, \textit{fault injection attacks} with the idea to introduce glitches in the hardware in an attempt to stop the accelerator or extract information \cite{8342177} and \textit{denial of Service attacks} with the idea  to disable the use of the accelerator, for example, by invalidating the shared DRAM with a row hammers \cite{Weissman2020} attack. For a discussion on these attacks, we remained the reader to \cite{https://doi.org/10.48550/arxiv.2209.11158}, a complete survey on the topic.
Fortunately, these attacks are known to the research community, and several countermeasures have already been implemented. Amazon AWS, for example, implements a bitstream antivirus \cite{https://doi.org/10.48550/arxiv.2209.11158} that scans the sent bitstream for harmful structures before programming the \gls{fpga}. Also, it only allows the loading of bitstreams generated with its execution stream. Additionally, most recent \glspl{fpga}, such as those from AMD Xilinx, include isolated partial reconfiguration capabilities that allow for independent partitioning and scheduling of \gls{fpga} resources and configure secure access \cite{10.1145/3560834.3563832}. However, like every security countermeasure, these solutions negatively impact the power consumption of the data center.
}

As we have often pointed out, the high power consumption of accelerators, along with their under-utilization, can impose high operating costs~\cite{Zhang11}. This sentence explains why virtualization is so relevant. \textbf{\gls{avm} placement} problem is well known and discussed in the literature, but not many papers extend the problem by considering \glspl{fpga}. Yarahmadi et al.~\cite{9544500} introduce a genetic algorithm-based \gls{vm} placement method capable of responding to these needs. The goal is to minimize the postponement of the requests while staying energy-efficient and compensate for the enormous exploration space created by limiting the number of chromosomes as a function of the number of \glspl{vm}. Zhang et al.~\cite{8644591} present a similar work with the difference that, in this case, they provide an advanced energy model of the data center.

Correlated to the placement problem, we always find \textbf{Accelerated job scheduling} problems. Scheduling on \gls{fpga} systems can be considered a classical \gls{rcsp}. We know the classical \gls{rcsp} problem as \gls{avm} placement is a strongly NP-hard problem. Thus, the \gls{fpga} community dedicates much attention to the design of heuristics~\cite{9217815}. Many research groups have approached this problem under different aspects~\cite{7082811}~\cite{8742331}~\cite{7411304}. Dhar et al.~\cite{9661327} propose a methodology for scheduling heterogeneous tasks across an \gls{fpga} considering the possibility of partial reconfiguration and sharing of resources. This methodology reduces the resource fragmentation and can optimize spaces and leakage power. Ting Loke and Yang Koay~\cite{7929526} instead approach the problem from an innovative point of view. Their idea consists in scheduling the tasks considering the relative deadline for each one. If a task has a lot of time available, they increase its latency by applying \gls{dfs} optimizations and clock gating at run time. In this way, they obtain an adaptive model that regulates energy optimization according to the time available and the type of task. 

We cannot conclude our discussion without talking about the main functionality that \glspl{fpga} offer; the \textbf{reconfiguration}. The ability to reconfigure the hardware allows the data center to implement different energy policies during computation and to test optimizations and solutions that would otherwise be impossible without updating the entire hardware. This feature is not free, and we must consider the times introduced by the process. Updating the bitstream of a single \gls{fpga} can take several seconds. Organizing the data center with a distributed \gls{fpga} placement and the use of \gls{fpga} overlays greatly reduces this overhead, guaranteeing rapid updating of entire clusters contemporary. Table~\ref{tab:infraSummary} summarise the optimization techniques and the main works presented in this subsection.

\vspace{4pt}
{\sc OVERALL IMPACTS CONSIDERATION: }\textit{The techniques presented in this subsection aim to maximize the use of \glspl{fpga} in the data center to enhance investment. This procedure increases the energy consumption related to the individual \glspl{fpga}, even if limited, to reduce the workload on the more traditional CPU-based computing nodes and, therefore, their consumption. \gls{fpga} resource virtualization harms network consumption but positively on storage consumption, allowing to share of memory resources between \glspl{fpga} across the network and maximizing their use. We can do the same reasoning with accelerated virtual machine placement, where the aim is to increase the utilization of the single node and avoid having to switch on more nodes to manage a specific workload. Finally, we discuss the fundamental feature of \glspl{fpga} reconfiguration. The reconfiguration acts on every aspect of the data center, allowing the deployment of \glspl{fpga} into servers, network devices, and control systems. The reconfiguration keeps the data center updated for future challenges by guaranteeing flexibility and maintainability.} \rev{We consider virtualization or resource-sharing techniques on \glspl{fpga} to be vital for the development of efficient and sustainable data centers.}

\begin{table}[!t]
\renewcommand{\arraystretch}{1.25}
\centering
\caption{Infrastructure Level - Power optimization techniques summary}
\label{tab:infraSummary}
\vspace{-6pt}
\begin{tabular}{@{}m{2.5cm} m{2.5cm} c c @{}}
\toprule
 \multicolumn{1}{c}{\textbf{Techniques}} &
 \multicolumn{1}{c}{\textbf{Main works}} &
 \multicolumn{2}{c}{\textbf{Target power}} \\
   \cline{3-4}
& & \textbf{Dynamic} & \textbf{Leakage} \\
\midrule 
 \gls{fpga} resource virtualization & \cite{8672857} \cite{1568522} \cite{6239797} \cite{3506713} \cite{7082811}  \cite{Hayden16} \cite{Huigui16} \cite{3287324.3287475}  & \ding{55} & \ding{51}\\
 \rev{Accelerated virtual machine} & \cite{9544500} \cite{8644591}& \ding{55} & \ding{51}\\
 Accelerated job scheduling &  \cite{9217815} \cite{7082811} \cite{9661327} \cite{8742331} \cite{7929526} \cite{7411304} \cite{6339197}& \ding{51} & \ding{51}\\
 Reconfiguration & \cite{Putnam14} \cite{Putnam15} & \ding{51} & \ding{51}\\
\bottomrule
\end{tabular} 
\end{table}

\section{Power Optimization Trends}
\label{sec:powerTrends}
Data center systems are subject to constraints such as \gls{tco}, \gls{pue}, performance, resilience, modularity, and scalability that reduce the degree of freedom on the possibility. From the Bobda et al.~\cite{3506713} analysis, all service providers prefer to have \glspl{fpga} in a distributed architecture that guarantees maximum flexibility and scalability to the system. Integrating off-the-shelf products such as existing data center upgrades is rarely possible. The form factor, the cooling system, and the constraints on the consumption of the single server are the main reasons why this integration does not take place. This problem leads to investing in custom solutions, which, in addition to being specific for a single environment, are also better designed from an energy point of view, allowing the board voltages to be monitored and managed more precisely.

Optimization techniques widely used today are power gating, approximate computing, \gls{avm} placement, and accelerated job scheduling. The reasons are many. Power gating is a simple technique that does not require calibration systems and does not introduce further problems as it does for \gls{dvs}. Since \glspl{fpga} do not have specific hardware for floating-point computation, optimizations on the type of data used can almost always be found by preferring solutions where the calculation is in fixed-point. This practice is common in data centers where the primary workload is statistical data analysis. Machine learning algorithms are resilient to data approximation techniques, guaranteeing the same results. \gls{avm} placement and accelerated job scheduling are extensions of the already present and widely used \gls{vm} management algorithms in the data center. These techniques provide great flexibility in managing the workload and servers available, which is essential for the performance, resilience, and modularity that a data center must have.

\section{Potential Future Innovations}
\label{sec:futureInnovation}
We identify areas of potential novelty by considering the optimization techniques used today and comparing them with the needs presented in the woks summarized into Tables \ref{tab:technologySummary}, \ref{tab:rtlSummary}, \ref{tab:allSummary}, \ref{tab:depSummary}, and \ref{tab:infraSummary}.
\\
\\
\textbf{Power models: }The most commonly used energy models do not consider the device temperature. Works such as the one presented by Khaleghi et al.~\cite{8988683} show us how seeing the thermal profile of the chip allows more studied optimizations. The Thermodynamics computing model~\cite{arXiv:1911.01968} is an excellent replacement for today's model. Unifying the entropy of Shannon with the entropy in thermodynamics and with the metrics of Kolmogorov would allow us to estimate the energy and thermal consumptions from the drafting of the algorithm alone without having to simulate the obtained solution from time to time. Some work exists in the literature but is still very limited and hard to use in real applications~\cite{DBLP:journals/corr/cs-IT-0410002}. This model would then find a particular application in \glspl{fpga} where high-level code directly impacts the generation of the hardware, allowing optimization of energy and temperature of the chip contemporary. \rev{Also, there is a need for new metrics that consider the type of energy consumed, whether it comes from renewable sources or fossil sources. Section 3 proposes an example, but the search field is still open. Finally, data center providers could consider using energy consumption as a unit of measurement for the utilization of their resources as they already do with time units. So, for example, setting a consumption limit as a constraint for the user.}
\\
\\
\textbf{Type of boards: }As discussed in Section \ref{sec:powerTrends}, the numerous data center constraints prevent the direct integration of off-the-shelf products. A fundamental step in the diffusion of \glspl{fpga} in this sector is the development of modular boards that can be assembled according to specific needs~\cite{3506713}, thus reducing the costs of experimenting and designing custom solutions. The possibility of adding exact control modules for energy would allow monitoring of the consumption of individual servers with greater granularity and smoothly implement optimization techniques on a physical level. \rev{This is also what the data center providers need to respect the code of conduct (points \ding{204}, \ding{205}, and \ding{206}).}
\\
\\
\textbf{Adaptive systems: }Introducing new hardware into an architecture leads to management challenges. It is not always possible to individually optimize each accelerator for a specific workload. Adaptive systems on \glspl{fpga} able to adopt the energy profile best suited to a given workflow would greatly help to manage this technology and its diffusion. Systems of this type could strongly limit energy consumption by identifying invisible configurations at design time, even going so far as to self-reconfigure to work better.
\\
\\
\textbf{Management tools: }The most famous platforms, such as Openstack~\cite{8711829} and Kubernetes~\cite{8968907}, which instantiate, and reconfigure data center resources in real-time, can effectively optimize energy consumption but do not have support for \glspl{fpga}. Extending these tools with \gls{fpga} support is, of course, a way to ensure the continued use of \glspl{fpga} in data centers even in the long term. In particular, the new possibility should be to virtualize these new resources, monitor their consumption and possible configuration errors (as appears for Microsoft's Catapult II project~\cite{Putnam15}), and easily reconfigure them remotely. All act with different granularity at the single server, rack, or cluster level.

\subsection{Green data distribution and challenges}
Recently Google introduced the concept of \textbf{Carbon-Aware Computing}~\cite{9770383}, which exploits flexibility in when, where, and how computing takes place to reduce carbon emissions. The idea is not new. Some computing has flexibility in when it can run, like processing videos, feature extraction and training large-scale machine learning models, simulation pipelines, and many other latency-tolerant workloads. So, they shape the load profile over 24 hours to balance the absorbed pick energy. Some computing jobs have flexibility in where they can run, like user-facing services that can be geographically rebalanced or resourced (common for Microsoft's applications, Facebook, and Twitter messaging applications) to a greener data center like the one with \gls{fpga} support. Today there are two deployed systems Microsoft's Carbon Aware Kubernetes~\cite{kubernetes20} and Google's Carbon-Intelligent computing platform~\cite{google21}. The challenges, in this case, consist of understanding which types of workloads enjoy the properties presented, ensuring that spatially flexible load ends up in the right location, and how load shaping affects \rev{$CO_2$} emission.

\section{Conclusion}
\label{sec:conclusion}
The article exposes today's state-of-the-art energy optimization techniques involving \glspl{fpga} in data centers. These techniques improve and enhance the benefit given by \glspl{fpga} with a view to sustainable computation. The article focuses on metrics, monitoring power consumption models, existing \glspl{fpga} power optimization techniques, and open challenges we must achieve in energy efficiency. Data center designers must understand this information well to prioritize energy consumption during design. 
Optimization techniques widely used today are power gating, approximate computing, \rev{\gls{fpga} resource virtualization}, \gls{avm} placement, and accelerated job scheduling. These techniques push the utilization of \glspl{fpga} to the maximum, their performance, and their energy efficiency while also giving great flexibility in managing the workload and servers available, which is essential for the performance, resilience, and modularity that a data center must-have.
We believe that \glspl{fpga}, with the proper observations and thanks to their low latency, high throughput, and energy efficiency, can be a complementary alternative to \glspl{cpu} and \glspl{gpu}. Especially, when the principal workloads are statistical data analysis, AI, computer vision, and security. The evolution of this technology, in the short term, will determine its long-term use in data centers. Many obstacles are still present. The main one is the lack of support for \glspl{fpga} in the asset management and data center monitoring tools. Besides, there is no comprehensive software stack to deploy them on the cloud. The extension of software such as OpenStack and Kubernetes to \glspl{fpga} is what cloud service providers are looking for to consider \glspl{fpga} an attractive product. Much has been done and will be done in the future to make \glspl{fpga} more attractive. Just think that \glspl{fpga} was born to facilitate the design of \glspl{asic}, and today we find them in the largest data centers in the world. In the future, we will probably see the birth of two types of products, one focused on performance and one focused on consumption and therefore to greater diversity and applicability of this technology, for sure if we will consider the need for modular and easily customizable boards.

\section*{Acknowledgments}

This work was supported by the Italian Ministry of University and 
Research (MUR) and the European Union (EU) under the PON/REACT project and by the Horizon 2020 EU Research \& Innovation Programme under grant agreement No. 957269 (EVEREST project).


\bibliographystyle{IEEEtran}
\bibliography{main}

\begin{IEEEbiography}[{\includegraphics[width=1in,height=1.25in,clip,keepaspectratio]{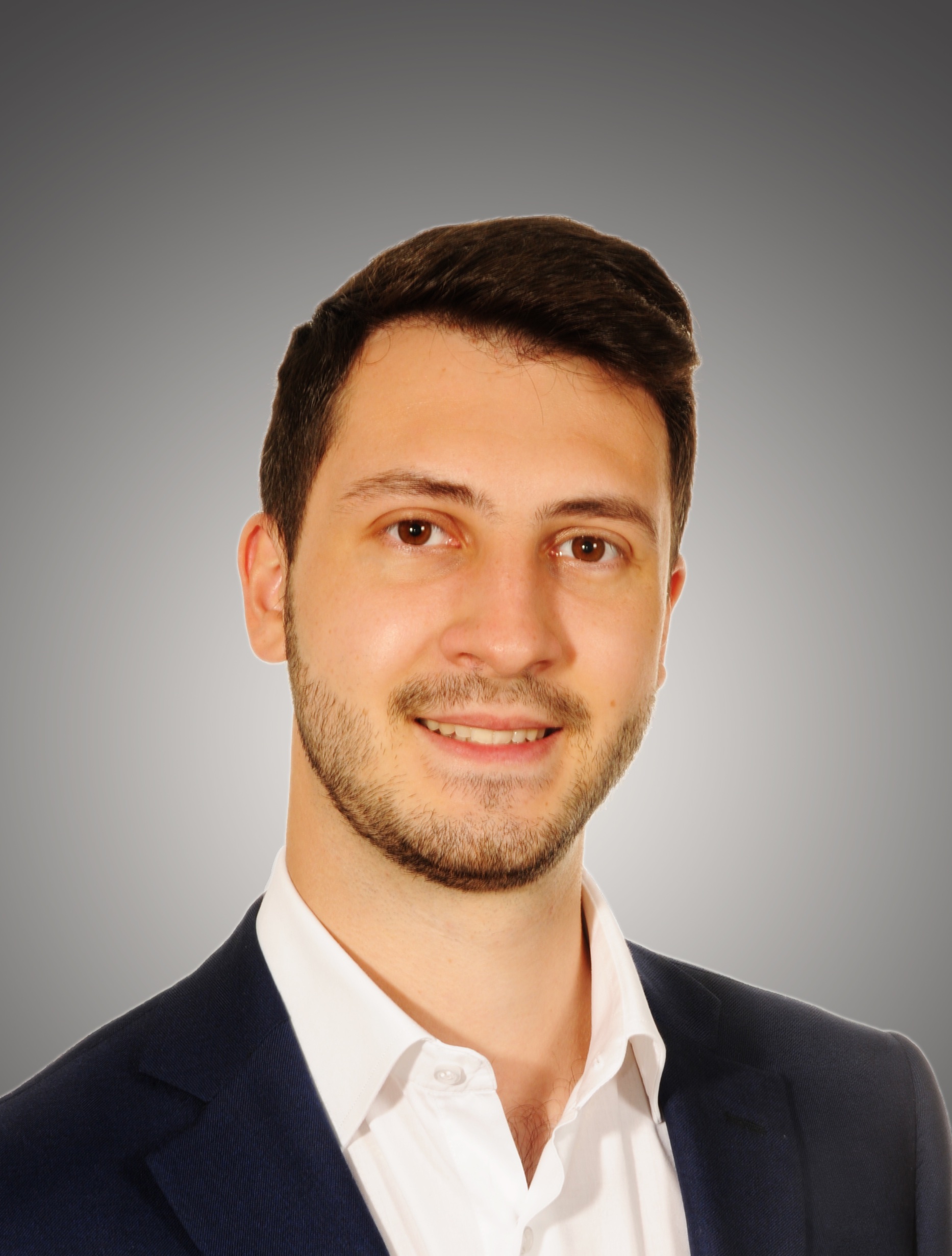}}]{Mattia Tibaldi} is a Ph.D. candidate at Politecnico di Milano. He received a Master's degree in Computer Science and Engineering from the same university in 2021. His research interests include reconfigurable and adaptive systems for cloud applications and data centers, security-aware design methodologies, and sustainable computing.
\end{IEEEbiography}

\begin{IEEEbiography}[{\includegraphics[width=1in,height=1.25in,clip,keepaspectratio]{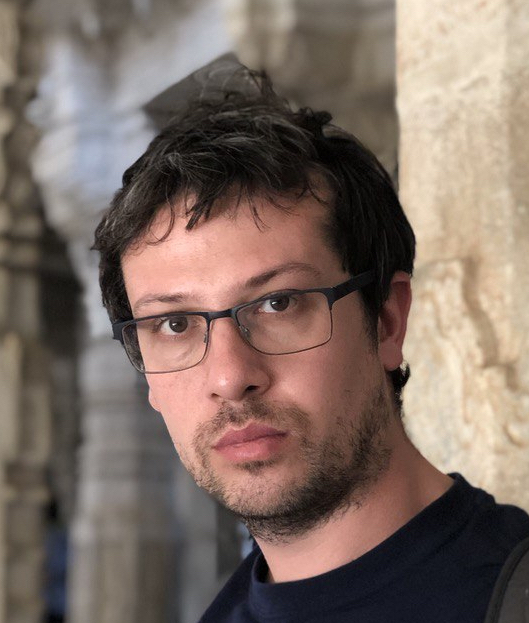}}]{Christian Pilato} is an Associate Professor at Politecnico di Milano. He was a Post-doc Research Scientist at Columbia University (2013-2016) and Università della Svizzera italiana (2016-2018), and an Assistant Professor at Politecnico di Milano (2018-2023). He was also a Visiting Researcher at New York University, TU Delft, and Chalmers University of Technology. He has a Ph.D. in Information Technology from Politecnico di Milano (2011). His research interests include high-level synthesis, reconfigurable systems, and system-on-chip architectures, focusing on memory and security aspects. He served as program chair of EUC 2014 and ICCD 2022 and is currently serving on the program committees of many conferences on EDA, CAD, embedded systems, and reconfigurable architectures (like DAC, ICCAD, DATE, CASES, FPL, ICCD, etc.) He is a Senior Member of IEEE and ACM, and a Member of HiPEAC.	
\end{IEEEbiography}

\end{document}